%% file: ms.tex
\documentclass[12pt,preprint]{aastex}

\def\msun{$M_{\odot}$}

\def\hbeta{H$\beta$ }
\def\hgamma{H$\gamma$ }

\def\Te{T_{\rm eff}}
\def\Topt{T_{\rm opt}}
\def\Tuv{T_{\rm UV}}
\def\Tv{T_V}
\def\logg{\log g}
\def\gta{\lower 0.5ex\hbox{$\buildrel > \over \sim\ $}} 
\def\lta{\lower 0.5ex\hbox{$\buildrel < \over \sim\ $}} 

\tighten

\begin{document}

\title{A COMPARATIVE STUDY OF OPTICAL AND ULTRAVIOLET EFFECTIVE TEMPERATURES 
FOR DA WHITE DWARFS FROM THE {\it IUE} ARCHIVE}

\author{C.-P. Lajoie}
\affil{Department of Physics and Astronomy, McMaster University,
  Hamilton, Ontario, Canada, L8S 4M1}
\author{P. Bergeron}
\affil{D\'epartement de Physique, Universit\'e de Montr\'eal, C.P.~6128, 
Succ.~Centre-Ville, Montr\'eal, Qu\'ebec, Canada, H3C 3J7.}
\email{lajoiec@physics.mcmaster.ca, bergeron@astro.umontreal.ca}

\begin{abstract}
We present a comparative study of effective temperatures determined
from the hydrogen Balmer lines and from the UV energy distribution for
140 DA white dwarfs drawn from the {\it IUE} archive. Our results
indicate that the optical and UV temperatures of the majority of stars
below $\Te\sim40,000$~K and within $\sim 75$ pc are in fairly good
agreement given the uncertainties. At higher temperatures and/or
larger distances, however, significant discrepancies are observed.
Several mechanisms are investigated to account for these discrepancies
including the effect of interstellar reddening, the presence of metals
in the photosphere, and the existence of unresolved binary white
dwarfs. The results of our analysis reveal that wavelength-dependent
extinction is the most natural explanation for the observed
temperature differences. We also attempt to predict the differences in
optical and UV temperatures expected from unresolved degenerate
binaries by performing an exhaustive simulation of composite model
spectra. In light of these simulations, we then discuss some
known double degenerates and identify new binary candidates by
restricting our analysis to stars located within 75 pc where the
effect of interstellar reddening is significantly reduced.

\end{abstract}



\section{INTRODUCTION}

White dwarf stars are known to exist in two varieties: DA and non-DA
stars. This spectral classification is primarily based on the presence
(DA) or absence (non-DA) of hydrogen lines in their optical spectra.
More precisely, non-DA stars have atmospheres mostly composed of
helium but sometimes show weak hydrogen lines and prominent lines of
heavier elements in the optical \citep[see, e.g.,][]{wesemael93}. The
optical spectra of DA stars, on the other hand, are completely
dominated by broad hydrogen absorption lines, while the ultraviolet
(UV) part of their spectrum may also reveal the presence of weak
metallic lines, although these lines tend to appear only at high
effective temperatures ($\Te\ \gta 40,000$ K). From a theoretical
point of view, the hydrogen atom is now fairly well understood, and
the atmospheres of DA stars can be modeled to a high level of
complexity. Fundamental parameters such as effective temperature and
surface gravity can therefore be derived quite accurately for large
samples of DA stars over a wide range of atmospheric parameters.  The
mass of these stars can also be obtained from detailed evolutionary
models \citep[e.g.,][]{wood95}. For instance, early pioneering work by
\citet{BSL}, and more recently by \citet[][hereafter LBH05]{liebert2005}, have used 
such model atmospheres to derive fundamental parameters of large
samples of DA stars using this so-called spectroscopic
method. Spectral analyses of different parts of the spectrum, however,
and in particular a comparison of the atmospheric parameters obtained
from other spectral regions could in principle help reveal
interesting objects and/or possible caveats in the modeling of DA star
atmospheres.

With the advent of UV astronomy, and in particular with the launch of
the {\it International Ultraviolet Explorer} ({\it IUE}) space
telescope in 1978, studies of this part of the spectrum have become
possible. During its eighteen-year mission, {\it IUE} observed
more than 320 degenerate stars at both low or high
resolutions. Although the {\it IUE} spectral coverage does not fully
include the Ly$\alpha$ line, the slope of the $\sim2000$~\AA\ wide
energy distribution can be used successfully to measure the effective
temperature. This new spectral window thus offers an excellent
opportunity to verify the internal consistency of fundamental
parameters obtained from different parts of the energy
distribution. Unfortunately, only a few such studies exist in the
literature. For instance, \citet{barstow2001,barstow2003a} relied on
\textit{FUSE} data to estimate $\Te$ from the Lyman series, and
they found significant discrepancies when compared with temperatures
obtained from the Balmer line profiles in DA stars hotter than
$\Te\sim50,000$~K. Although most DA stars with $\Te\ \gta 50,000$~K
are now known to contain small traces of metals in their atmosphere,
the authors attributed the observed discrepancies to deficiencies in
the detailed physics incorporated in the model atmosphere
calculations. The effect of metals on the determination of effective
temperatures using the UV part of the spectrum has not been studied
systematically, but some authors
\citep{lanz1995,barstow2003a} have suggested that it might lead to
an overestimate of $\Te$ when compared to the optical determinations.

Assuming that the model atmospheres are accurate enough, discrepancies
in the determination of the atmospheric parameters using different
parts of the energy distribution might also reveal the presence of
unresolved double degenerate (DD) binaries. This is appealing because
binary white dwarfs are thought to be responsible for type Ia
supernova explosions (SN Ia). In this scenario, two massive white
dwarfs merge, resulting in an object with a mass exceeding the
Chandrasekhar limit of $\sim1.4$~\msun. Carbon and oxygen burning
in degenerate conditions is then believed to lead to a thermonuclear
runaway that completely disrupts the star. The search for these DDs
has become a hot topic of research over the last few years, although
observations of the bulk of this population has remained elusive,
mostly because of their small expected angular separations. Radial
velocity searches \citep[e.g.,][]{saffer98,mm1999,mmm2000} have
revealed the presence of a few DDs with total masses close to the
Chandrasekhar mass limit, but the numbers are still too
low. Furthermore, most of the few known DDs have periods too long for
them to merge within a Hubble time. All in all, there simply seems to
be too few of them to account for the observed rate of SN Ia
\citep{robinson,bragaglia1990,foss1991,saffer98}. But there is
hope: large surveys such as the ESO \textit{Supernova Ia Progenitor Survey}
\citep[][SPY]{napi2001} are now actively looking for these DD systems and
should yield interesting results in the near future.

With these ideas in mind, we conducted a comparative study of effective
temperatures determined from optical and UV observations for DA stars
drawn from the {\it IUE} archive with the goal of identifying objects
showing significant discrepancies in their temperature estimates. We
describe respectively in \S~\ref{sect:observ} and \S~\ref{sect:models}
the observational material and theoretical framework used in our
analysis. The effective temperature and surface gravity estimates are
then presented in \S~\ref{sect:results} and discussed in
\S~\ref{sect:discuss} where we show that most hot DA stars in our
sample show significant discrepancies in their effective temperature
determinations. Several possible sources of contamination are
investigated, including interstellar reddening, the presence of heavy
elements in the photospheric regions, and binarity. On the basis of
our results, we then discuss in \S~\ref{sect:objects} several
interesting objects, including double degenerate candidates. Our
conclusions follow in \S~\ref{sect:concl}.


\section{OBSERVATIONS}\label{sect:observ}

Our main goal was to obtain the optical counterpart of all the DA
stars observed by {\it IUE} during its 18-year mission.  The {\it IUE}
low-dispersion spectra used in this study come from the new reduction
procedure of \citet[][hereafter HBB03]{holberg2003}, which follows the
prescription of \citet{massa} for the correction of
residual temporal and thermal effects, as well as absolute flux
calibration. In fact, it is because this procedure is optimized
for low-dispersion spectra of hot continuum sources that HBB03 have
applied it to the complete sample of white dwarf stars observed by
{\it IUE}. Given these corrections, along with the coaddition of multiple
observations of the same object, this new reduction allows a
significant gain in the signal-to-noise ratio (3\% of the signal for
optimally exposed spectra). The spectral coverage for the {\it IUE} spectra
is about $\lambda\lambda$1150-1970 (SWR camera) and
$\lambda\lambda$1850-3150 (LWR camera) at a resolution of $\sim$6 \AA\ 
and $\sim$0.2 \AA\ for the low- and high-dispersion modes, respectively. When
combined, these two data sets cover a spectral range of about 2000 \AA.
Note that for some stars, only one of the two spectral regions is
available.  For more details on the cameras and the different setups
used by {\it IUE}, see HBB03.

Among the $\sim$320 degenerates observed by {\it IUE}, more than half
are non-DA stars such as DB, DO, DP, DQ, DZ, or DC stars. Among the
remaining DA stars, several have a bad UV spectrum that cannot be used
in our analysis (WD 0104$-$464, 0216$+$143, 0250$-$026, 0517$+$307,
0518$-$105, 0531$-$022, 0646$-$253, 1053$-$550, 1159$+$803,
1413$+$231, 2116$+$736, 2205$+$250, and 2237$+$819). In addition, the
{\it IUE} spectrum of WD 0935$-$371, which is a DA$+$DQ binary system,
is actually that of the DQ star. Also, WD 0252$-$055, 0308$+$096,
0353$+$284, 0429$+$176, 1347$-$129, 1550$+$130, and 2110$+$300, all
classified DA stars in HBB03, are found in binary systems with a
bright companion that dominates or contaminates substantially the
optical flux. These white dwarfs have thus been eliminated from our
sample.

The optical spectra for the remaining DA stars have been obtained
over several observing runs using the Steward Observatory 2.3 m
telescope equipped with the Boller \& Chivens spectrograph. The
spectral coverage is about $\lambda\lambda$3100-5300,
thus covering \hbeta to H9 at an intermediate resolution of $\sim$6
\AA\ (FWHM). Objects south of declination $\sim-30$ degrees were excluded
from our sample, although a few spectra in the Southern hemisphere
were available from the analysis of \citet{bragaglia1995} or provided
by C.~Moran (1999, private communication); these have a spectral
resolution ranging from 3 to 9 \AA\ (FWHM). Among the DA stars
selected in our original sample, WD 0109$-$264, 1121$+$145,
1544$+$009, and 2333$-$002 turned out to be subdwarf stars, while WD
1055$-$072, previously classified DA7 in
\citet{mccook99}, is actually a featureless DC star. WD 0945$+$245
(DA+DX), 1015$+$014, and 1031$+$234 are magnetic white dwarfs, which
cannot be easily analyzed within our theoretical framework, while WD
0950$+$139 shows strong emission lines produced by the faint planetary
nebula. Also, WD 2246$+$066 (HS 2246$+$0640) is too hot ($\Te\sim
140,000$~K according to our own estimate) and the Balmer lines are too
weak to be analyzed with sufficient accuracy in our study.  Finally,
WD 1735$-$318 was too faint ($V=18.1$) to be observed at Steward,
while no references could be found in \citet{mccook99} for WD
0511$-$230.  All in all, we end up with a sample of 140 DA stars for
which both optical and UV spectra are available.  The optical spectra
of this sample are displayed in Figures
\ref{fg:f1} and \ref{fg:f2} in order of decreasing effective temperature, 
as determined in the next section.


\section{MODEL ATMOSPHERES AND FITTING TECHNIQUE}\label{sect:models}

The model atmospheres used in this analysis are described at length in
LBH05 and references therein.
The DA stars drawn from the {\it IUE} archive cover a wide range of
effective temperatures including hot stars where non-local
thermodynamic equilibrium (NLTE) effects are important ($\Te\
\gta40,000$~K), and cooler stars where energy transport by convection
dominates over radiative transport ($\Te\ \lta15,000$~K). Our model
grid allows for both NLTE effects as well as energy transport by convection
following the ML2/$\alpha=0.6$ prescription of the mixing-length
theory (see \citealt{B95}). The theoretical spectra are calculated
within the occupation formalism of \citet{HM88}, which provides a
detailed treatment of the level populations in the presence of
perturbations from neighboring particles, as well as a consistent
description of bound-bound and bound-free opacities. Our pure hydrogen
model grid covers a range of effective temperature between $\Te=1500$
and 140,000 K by steps of 500~K for $\Te<20,000$~K, and by steps of
5000~K above, and a range of $\logg$ between 6.5 and 9.5 by
steps of 0.5 dex, with additional models at $\logg=7.75$ and 8.25.

The atmospheric parameters, $\Te$ and $\logg$, are first determined
from the optical spectra using the so-called spectroscopic method
\citep[see, e.g.][]{BSL}, which relies on a comparison between
synthetic and observed normalized profiles of the hydrogen Balmer
lines. LBH05 have recently improved upon this method by using
pseudo-Gaussian profiles or theoretical spectra to better define the
continuum of each line (see LBH05 for more details). Given the fact
that these lines are very sensitive to both $\Te$ and $\logg$, the
normalized profiles for \hbeta to H8 are compared to model spectra,
convolved with the appropriate Gaussian instrumental profile, using
the nonlinear least-squares method of Levenberg-Marquardt
\citep{press86}.  Note that this $\chi^2$-minimization procedure uses
all Balmer lines simultaneously to determine the atmospheric
parameters.  For cases where the red portion of the spectrum is
contaminated by an unresolved main-sequence companion, we neglect
\hbeta (WD 1026$+$002 and 1314$+$293) and, if necessary,
\hgamma as well (WD 0131$-$163, 0824$+$288, 1631$+$781, and 1636+351). 
Because of some irregularities in the line wings, \hbeta has also been
omitted in the fits of WD 0410$+$117, 1650$+$724, and 1749$+$717.

For the UV portion of the spectrum, we determine $\Te$ in two
independent ways. The first one relies on the slope of the UV energy
distribution, which, as shown for instance 
by \citet[][Fig.~10]{B95}, is not very sensitive to
surface gravity. For any
assumed value of $\logg$, it is always possible to adjust the
effective temperature to yield an equally good fit to the UV
observations. To overcome this problem, we fix $\logg$ to the optical
value and determine $\Te$ using the slope of the UV energy
distribution. Our fitting technique relies again on the nonlinear
least-squares method of Levenberg-Marquardt. Here, the observed fluxes
$f_\nu$ are compared to model Eddington fluxes $H_\nu$ using the
relation

\begin{equation}
f_\nu=4\pi(R/D)^2 H_\nu(T_{\rm eff},\log g)
\end{equation}

\noindent where only $\Te$ and the solid angle $\pi(R/D)^2$ are considered free
parameters.  $R/D$ is the ratio of the radius of the star to its
distance from Earth. Note that this method for determining $\Te$ is
only sensitive to relative fluxes, and it is therefore not subject to
any error in the absolute flux calibration. However, any effect that
may alter the {\it shape} of the UV energy distribution (e.g., interstellar
reddening) might affect our $\Te$ estimates significantly. We will
discuss some of these effects in \S~\ref{sect:discuss}.

We obtain a second independent estimate of $\Te$ by normalizing the UV
spectra to the measured $V$ magnitude of each star (see also
\citealt{finley90}). To do so, we first convert $V$ into a mean flux
using the relation

\begin{equation}
V=-2.5\log{f_\lambda^V} + C_V
\end{equation}

\noindent where $f_{\lambda}^V$ is the average flux in the $V$ bandpass, and
$C_V=-21.0607$ is the flux constant for photon counting devices
obtained by
\citet{holberg06} using new spectroscopic observations for Vega. 
We then divide each point of the observed and synthetic spectra by
their respective flux at $V$ and use the same minimization technique
as above to estimate $\Te$. With this approach, $V$ is in fact used as a
reference point around which synthetic spectra can be rotated until
the UV flux is matched. This method is thus extremely sensitive to the absolute
flux calibration scale, as well as to the accuracy of the $V$
measurements. If for any reason the absolute fluxes are underestimated
({\it IUE} had only a 3\arcsec-diameter circular aperture, which could
result in 50\% or less light transmission in some cases) or if the $V$
magnitudes have large uncertainties, this $V$-normalization method
could lead to fairly bad estimates of the effective temperature.

Examples of both methods --- UV-slope and $V$-normalization --- for
determining $\Te$ from the UV energy distributions are illustrated in
Figure \ref{fg:f3}. We thus end up with three different estimates of
$\Te$ for most of the 140 DA stars in our sample (4 objects in our
sample do not have $V$ measurements). These are compared in the next
section. Interesting objects such as the unresolved double degenerate
systems WD 0101$+$048 \citep{mmm2000}, 0135$-$052 \citep{saffer88},
and 1022$+$050 \citep{mm1999} are included in our sample.  Any discrepancy
between their $\Te$ estimates could help confirm their binary status
and validate our comparative approach as an efficient way to identify
double degenerates.


\section{RESULTS}\label{sect:results}

Effective temperatures derived from the optical spectrum
($\Topt$), the slope of the UV energy distribution ($\Tuv$), and
the $V$-normalization method ($\Tv$) for each object in our sample
are reported in Table 1, together with the $V$ magnitude, the surface
gravity obtained from the optical solution ($\logg$), the mass (in
solar masses), and the photometric distance ($D$). The $V$ magnitudes
are taken from the online version of the Villanova White Dwarf
Catalog\footnote{http://www.astronomy.villanova.edu/WDCatalog/index.html}
and references therein, with the exception of WD 0421$+$740,
1650$+$724, 1827$+$778, and 2207$-$303, for which no measurements were
available.  Masses and radii are determined from the evolutionary
models of \citet{wood95} with carbon-core compositions, helium layers
of $q({\rm He})\equiv M_{\rm He}/M_{\star}=10^{-2}$, and thick
hydrogen layers of $q({\rm H})=10^{-4}$. Photometric distances are
obtained from the distance modulus, which is given by the relation

\begin{equation}
V-M_V=5\log_{10}D-5
\end{equation}

\noindent where $D$ is the distance to the star in parsecs, and $M_V$
is the absolute magnitude interpolated in the photometric tables of
\citet{holberg06}\footnote{see
http://www.astro.umontreal.ca/$\sim$bergeron/CoolingModels/} at the values
of $\Te$ and $\logg$ determined from the optical solution. When no $V$
magnitude is available, the distance is estimated from the solid angle
$\pi(R/D)^2$ obtained from the UV-slope method, assuming that the {\it
IUE} absolute fluxes are accurate.

Uncertainties for the optical solution are 1.2\% in $\Te$ and 0.038
dex in $\logg$ (see \S~2.4 of LBH05 for details). Uncertainties
in $\Tuv$ and $\Tv$ are typically much larger, and they
are estimated in the following way. We first calculate a standard
deviation $\sigma_0$:

\begin{equation}
\sigma_0^2=\frac{1}{N}\sum_{i=1}^{N} \Big[f_{\rm obs}(\lambda_i) - 
f_{\rm bf}(\lambda_i)\Big]^2
\end{equation}

\noindent 
where $f_{\rm obs}(\lambda_i)$ and $f_{\rm bf}(\lambda_i)$ correspond
respectively to the observed and the best-fit model fluxes at a given
wavelength $\lambda_i$, and $N$ is the number of data points in the UV
spectrum. We then change the effective temperature until the standard
deviation between the corresponding model spectrum and our best-fit
model equals $\sigma_0$, providing us with a rough estimate of the
1$\sigma$ uncertainty. Typically, errors in $\Tuv$ are much larger
than errors in $\Tv$ because of the small sensitivity of the UV slope
at high effective temperatures. For instance, the effective
temperature uncertainty of a star near $\Te\sim50,000$~K, can be as
large as the measurement itself, depending on the S/N ratio. On the
other hand, the normalization at $V$ provides much better leverage,
and small variations of $\Te$ rapidly decrease the quality of the
fit. The corresponding uncertainties are thus much smaller than with
the UV-slope method. However, the $V$-normalization method is also
very sensitive to the absolute {\it IUE} flux calibration, which is
difficult to estimate. Uncertainties with both methods are reported in Table 1.

Figure \ref{fg:f4} shows the mass distribution of all DA stars in our
sample as a function of the optical effective temperature
($\Topt$). Objects are distributed fairly uniformly in effective
temperature from $\sim7000$ K (log $\Te=3.85$ for WD 0135$-$052) up
to $\sim98,000$ K (log $\Te=5.0$ for WD 0615$+$655), and in stellar
mass in a range below $M\ \lta0.45$~\msun\ (for stars labeled in
Fig.~\ref{fg:f4}) up to the most massive objects like WD 0136$+$251
($M=1.22$~\msun) and 0346$-$011 (GD 50, $M=1.27$~\msun). The low-mass
degenerates are quite interesting from the point of view of binary
evolution since the main-sequence lifetime of their progenitor is
estimated to be much longer than the age of the Galaxy. These low-mass
stars are thus believed to be the result of binary evolution. We
address the question of how the existence of unresolved degenerate
binaries may affect our temperature estimates in \S~\ref{subsect:simul}.

The mass distribution of our sample is compared in Figure
\ref{fg:f5} to that obtained by LBH05 for the DA stars in the Palomar-Green (PG)
Survey. Only stars with $\Te>13,000$ K are considered here since at
lower temperatures, small amounts of helium are suspected to be brought
to the surface by the hydrogen convection zone, thus increasing the
atmospheric pressure and the inferred spectroscopic masses
\citep{bergeron90}. It is comforting to see that
both mass distributions compare favorably well despite the fact that
the {\it IUE} archive does not constitute in any manner a complete
statistical sample in magnitude or volume. We note in Figure
\ref{fg:f5}, however, a clear deficiency of low-mass stars in our
sample ($M<0.45$~\msun) with respect to the PG sample. These have
larger radii and should have been easily detected by {\it IUE} since
they are more luminous than their high-mass counterparts. This bias
against the presence of low-mass stars in our sample thus seems purely
coincidental.

Finally, and most interestingly, all known ZZ Ceti stars included in
our sample (WD 0133$-$116, 0858$+$363, 0921$+$354, 1116$+$026,
1236$-$495, 1307$+$354, 1425$-$811, 1559$+$369, 1647$+$591,
1855$+$338, 1935$+$276, and 2326$+$049) fall within the instability
strip as defined by \citet{alex2005,alex2006}, given the uncertainties
in $\Tuv$ and $\Tv$. No other objects are found within these limits,
except for WD 1022$+$050 (LP 550-52), which is a known double degenerate
\citep{mm1999}.


\section{DISCUSSION}\label{sect:discuss}

\subsection{Comparison of Optical and UV Temperatures}

The effective temperatures obtained from optical spectroscopy ($\Topt$
in Table 1) are compared with those obtained from the UV energy
distribution in Figure \ref{fg:f6} as a function of $\Topt$ and in
Figure \ref{fg:f7} as a function of distance.  The UV-slope ($\Tuv$ in
Table 1) and $V$-normalization ($\Tv$ in Table 1) methods are shown in
the upper and bottom panels, respectively.  Instead of showing
individual error bars in these plots, we use open circles to indicate
the stars for which optical and UV effective temperatures are
inconsistent (a few indicative errors bars are also displayed in
Fig.~\ref{fg:f6}).  Here, we adopt a conservative criterion and assume
that both estimates are inconsistent if the (absolute) difference in
temperature is larger than 1.6~$\sigma$ (a 90\% confidence level),
where $\sigma^2=\sigma^2(\Topt)+\sigma^2(\Tv\ {\rm or}\ \Tuv)$.

Figure \ref{fg:f6} shows that for stars below $\sim40,000$~K, the
optical and UV temperatures are in fairly good agreement
($\Delta\Te/\Topt\ \lta10\%$), while discrepancies up to 50\% are
observed for stars hotter than $\sim40,000$~K. In the case of the
UV-slope method, most of these large temperature differences are still
consistent within the uncertainties (filled circles) because of the
large errors associated with this method at high temperatures. In this
case, only a few objects have statistically significant temperature
differences. On the other hand, most of the large temperature
differences observed with the $V$-normalization method are
statistically inconsistent within uncertainties, suggesting that these
differences are real and that some systematic effects may affect the
temperature scale of hot DA stars.

Figure \ref{fg:f7} also reveals a tendency for larger discrepancies
with increasing distance. Furthermore, stars located within
$\sim75$~pc typically show a much better agreement between their three
temperature estimates, even though a few nearby ($D\ \lta75$~pc)
objects still exhibit significant temperature differences. This
suggests, once again, that some systematic effects may affect the
temperature scale of more distant stars and, to a lesser extent, some
nearby stars. Given the fact that hotter stars --- which are often the
most distant ones, as can be judged from the results of Table 1 --- and
distant stars show the most significant effective temperature
discrepancies, we first study the possibility that interstellar
reddening might be responsible for the trends shown in Figures
\ref{fg:f6} and \ref{fg:f7}.


\subsection{Interstellar Reddening}\label{subsect:reddening}

Selective extinction from interstellar gas and dust can alter the
shape of the UV energy distribution since it is more effective
at short wavelengths (shortward of optical wavelengths).  Following
the prescription of \citet{seaton79}, we have simulated the effects of
interstellar reddening on our optical and effective temperature
determinations. To do so, we first apply to a model spectrum at a
given effective temperature and surface gravity a wavelength-dependent
extinction with an assumed color excess $E(B-V)$.  We then fit this
redenned spectrum with our standard unreddened model grid.  We have
run this simulation for synthetic spectra at various effective
temperatures at $\logg=8$ and with different color excesses.  The
predictions for both the UV-slope and $V$-normalization methods are
displayed in Figure \ref{fg:f8} together with the results for our DA
sample. From these simulations, we see that interstellar reddening can
easily reproduce the overall differences in measured effective
temperatures, in the sense that a large color excess yields a large
temperature difference, and this effect is more pronounced at high
effective temperatures, a trend also observed in our
sample. This is not an unexpected result, of course,
since the UV part of the spectrum is more affected by reddening than
the optical regions, and the flattening of the energy distribution
results in an underestimation of the UV effective temperatures. The
results shown in Figure \ref{fg:f8} also reveal that the
$V$-normalization method is much more sensitive to reddening than the
UV-slope method. For instance, at $\Te=70,000$ K, the UV-slope method
requires a color excess of $E(B-V)\sim0.08$ in order to produce a
$\sim$30\% difference in temperatures, while the $V$-normalization
method requires a value of only $E(B-V)\sim0.03$ to yield the same
result. This is once again due to the fact that the $V$-normalization
method is much more sensitive to small changes in the absolute UV fluxes.

To further validate our hypothesis, we must demonstrate that the color
excesses required to account for the observed temperature
discrepancies are not totally unrealistic, and that they agree
reasonably well with observed values. We estimate these values for
several DA stars in our sample by using the relation of
\citet[][Fig.~1]{spitzer} between the hydrogen column densities and
the $E(B-V)$ color excess 

\begin{equation}
N({\rm H})\approx 5.9 \times 10^{21} E(B-V)
\hspace{0.5cm}{\rm [mag^{-1}\ cm^{-2}]}
\label{eq:spitzer}
\end{equation}

\noindent combined with measurements of the column density taken from the 
literature (see \citealt{wolff96}, \citealt{marsh1997},
\citealt{wolff98,wolff99}, \citealt{bannister}, and \citealt{good2004}).  
The above relation includes the contribution of H\textsc{i},
H\textsc{ii}, and H$_2$ and is consistent with X-ray and 21-cm
observations towards globular clusters. As an example, we show by open
circles in Figure \ref{fg:f8} several objects in our sample that have
been corrected for the effect of reddening following the method
described above.  We see that in most cases, column densities taken
from the literature lead to a much better agreement between
temperature estimates.  However, these same values of $E(B-V)$ do not
agree well with those inferred from the theoretical curves of the
UV-slope method. This is mostly because of the large uncertainties
inherent to this particular method, and the location of the stars
with respect to the reddening curves is to be interpreted with
caution in this case. On the other hand,
the values of $E(B-V)$ suggested by our theoretical simulations with
the V-normalization method are fairly well corroborated by most of the
measured values of $E(B-V)$. Moreover, the agreement in effective
temperatures is also largely improved. This strongly suggests that
reddening plays an important role in explaining the observed
discrepancies in our effective temperature estimates, at least for
distant stars.


\subsection{Presence of Metals in Hot DA Stars}\label{subsect:metals}

In this section, we investigate the effects produced by the presence
of metals in the atmosphere of hot DA white dwarfs on the optical and
UV energy distributions. Heavy elements have already been identified
in the spectrum of most DA stars hotter than $\sim40,000$~K from UV
and extreme-UV observations \citep[see, e.g.,][and references
therein]{vennes2006}. G191-B2B (WD 0501$+$527) is probably the most
observed such metallic DA star and ions such as He, C, N, O, Si, S, P,
Fe, and Ni have already been identified in its spectrum.  Since metals
block most of the flux at short wavelengths, the stellar flux is
redistributed at longer wavelengths, and their presence in the
atmospheres of hot DA stars could potentially affect the temperature
structures and energy distributions. Because the observed metal
abundances are significantly reduced in DA stars below $\sim40,000$~K,
radiative pressure has been naturally invoked as the mechanism
responsible for maintaining these heavy elements in the atmospheric
layers of hot white dwarf stars. This is precisely the effective
temperature above which we observe the largest temperature
discrepancies in Figures \ref{fg:f6} and \ref{fg:f7}.

To study this effect, we start from a metal-rich synthetic spectrum,
kindly provided to us by I. Hubeny (2005, private communication), and
which is representative of G191-B2B at $\Te=54,000$~K, $\logg=7.5$,
with the metal abundances given in \citet{barstow2003b}. This spectrum
is compared in Figure \ref{fg:f9} with that of a pure hydrogen
spectrum interpolated in our model grid for the same values of the
atmospheric parameters. Note that both spectra have been calculated in
a consistent manner using TLUSTY. From this comparison, we can see
that indeed, the presence of metals has increased significantly the
flux in the UV portion of the spectrum. We now fit the optical and UV
fluxes of this metal-rich spectrum in the same manner as described in
\S~\ref{sect:models} using our pure hydrogen model grid. The
effective temperature obtained from the Balmer line profiles is about
3000 K hotter than the true temperature of the model ($\Te=54,000$~K),
while both $\Tuv$ and $\Tv$ are about $\sim6500$ K hotter. Thus, the
difference between UV and optical temperatures expected in metal-rich
DA stars is {\it positive} when analyzed with pure hydrogen models,
that is, above the dotted lines in Figures \ref{fg:f6} and
\ref{fg:f7}. In other words, had we fitted the DA stars in our sample
with a grid of model atmospheres that include metals, we would have
found even larger temperature discrepancies than those already
observed here. This conclusion is similar to that reached by
\citet{lanz1995} and \citet{barstow1998,barstow2001,barstow2003a}, who found that $\Te$ is
overestimated in both optical and UV analyses when metals are present
in the atmosphere of hot DA stars. We are thus forced to conclude that
the temperature discrepancies observed in Figures \ref{fg:f6} and
\ref{fg:f7} cannot be easily explained by the presence of metals in
the hottest DA stars, or alternatively, by our use of pure hydrogen
model atmospheres.


\subsection{Unresolved Double Degenerates}\label{subsect:simul}

As discussed in the Introduction, one the goals of this project was to
search for unresolved double degenerate binaries by comparing
effective temperatures determined from optical and ultraviolet
spectra. As demonstrated by \citet{liebert91}, it is impossible to infer
the presence of such binary systems using the optical spectroscopic
technique alone. Indeed, the coaddition of synthetic spectra of two DA
stars with different values of $\Te$ and $\logg$ can be reproduced
almost perfectly by a single DA spectrum. Hence, double degenerates
would go totally unnoticed in an optical spectroscopic survey.

In this section, we extend the experiment of \citet{liebert91} by
including the information contained in the UV energy distribution.
Hence, we coadd the monochromatic fluxes of two synthetic spectra
with random values of effective temperatures and surface gravities,
properly weighted by their respective radius; because of the
mass-radius relation, low-mass white dwarfs contribute more to the
total flux than their more massive counterparts at the same
temperature.  We then fit these composite spectra and determine
optical and UV temperatures in the same manner as described in
\S~\ref{sect:models}. Since in the minimization procedure there 
are always a cool and a hot solution on each side of the maximum
strength reached by the Balmer lines, we always choose the solution
with the lowest $\chi^2$ value calculated from the difference between
the simulated and best fit model spectra normalized at 4600 \AA.  When
both fits are acceptable, which occurs when the temperatures of the
two components are comparable, we retain both solutions. The results
of our simulations up to $\Te=50,000$~K are displayed in Figure
\ref{fg:f10} for the UV-slope and $V$-normalization methods. At higher
temperatures, the hot component of the system always dominates the
total flux and the predicted differences in optical and UV
temperatures are too small to be measured efficiently with this
technique.

Our results reveal that many fitted composite spectra show a good
agreement between their three $\Te$ estimates. These binary systems
correspond to those for which the total flux is almost completely
dominated by the brightest component of the system, or to binaries
that have very similar atmospheric parameters. These double
degenerates would therefore go undetected with our comparative
approach. Second, the results of Figure \ref{fg:f10} indicate that
most DA stars in our sample with temperature discrepancies larger than
$\sim20$\% cannot be interpreted in terms of binarity alone, whatever
method is used. Given the uncertainties in our $\Te$ determinations,
double degenerates that could be detected by a comparison of optical
and UV temperatures would rather be found at effective temperatures
below $\Te\sim25,000$~K, or in regions where UV temperatures largely
exceed the optical temperatures with the UV-slope method.


\section{OBJECTS OF PARTICULAR INTEREST}\label{sect:objects}

Given the established dominant role of interstellar reddening in
explaining the observed temperature differences, we now concentrate on
nearby objects that are not significantly affected by reddening.  The
Sun is located in a region of the Galaxy where almost no dust and gas
are found. The dimension of this Local Bubble ranges from 65 to 150 pc
depending on the line of sight \citep{lallement}. Beyond this, a wall
of cold gas and dust with a typical H\textsc{i} column density of
$\sim3\times10^{19}$ cm$^2$ is observed. Thus, in order to minimize
the effect of interstellar reddening, we restrict our analysis to
objects located within 75 pc. Our results are presented in Figure
\ref{fg:f11} for both the UV-slope and 
$V$-normalization methods. Open symbols once again
represent objects for which the temperatures are inconsistent within
the uncertainties. These particular objects are labeled by their WD
number in the figure and they are further discussed below along with
other interesting objects.

{\it WD 0101$+$048 (G2-17), 0135$-$052 (L870-2)} --- WD 0101$+$048 
is a known binary star showing radial velocity variations
\citep{mmm2000}.  However, the orbital period is uncertain: $1.2$ or
$6.4$ days.  Furthermore, the spectroscopic mass of 0.77
\msun\ is inconsistent with the photometric mass of 0.37
\msun\ obtained by \citet{BLR} based on its trigonometric parallax
measurement. This discrepancy is likely due to the over-luminosity of
the binary system (compared to a single star): an over-luminous object
is then interpreted as a large-radius, less massive white
dwarf. The location of WD~0101+048 in both panels of Figure
\ref{fg:f10} cannot be easily accounted for by our DD simulations, however.
Furthermore, the optical and UV temperatures are consistent with both
the UV-slope and V-normalization methods. A similar conclusion can be
reached for WD~0135$-$052, another known double degenerate system
\citep{saffer88}, although in this case the temperature obtained from
the UV-slope method is inconsistent with the optical value
(Fig.~\ref{fg:f11}). The measured temperature differences for these
two stars are actually in the opposite sense from what is predicted by
our simulations. We believe that the problem may be related to the
inaccurate treatment of the UV opacity in our model atmosphere
calculations at low temperatures. In particular, \citet{kowalski06}
have shown that the red wing of the Ly$\alpha$ line of hydrogen can
become an important opacity source in the UV when perturbations of
hydrogen atoms by their interaction with molecular hydrogen are
properly taken into account. This opacity source is not included in
our model calculations and could affect our results.

{\it WD 0346$+$011 (GD 50)} --- GD 50 is a massive DA star
(1.27~\msun) showing unusual helium abundances \citep{vennes1996}.
Interestingly, one of most accepted formation scenarios for GD 50 is
the merger of two white dwarfs \citep{BSL,segretain,garcia97}.  The
high mass and high effective temperature of this object suggest that
GD 50 is relatively young, and a merger scenario would require this
object to be in a relatively dense stellar
environment. \citet{dobie06} have argued on the basis of astrometric
and spectroscopic data that GD 50 is actually a former member of the Pleiades
cluster. Both methods for estimating the UV temperature yield similar
values (Fig.~\ref{fg:f11} and Table 1). Only low column densities are
measured in the line of sight of GD 50 so the difference of optical
and UV temperatures cannot be explained in terms of
reddening. Moreover, no radial velocity variations have ever been reported
for this star. So given the particular nature of this object, it is
difficult to draw any firm conclusion about its possible binary character.

{\it WD 0348$+$339 (GD 52)} --- The difference in temperature observed
in Figure \ref{fg:f11} with the V-normalization method is only a
1.7~$\sigma$ result, and it is therefore not considered significant.

{\it WD 0509$-$007 (RE J0512$-$004)} --- Even though all three $\Te$
estimates agree well for this object ($\sim31,000$ K), its low mass of
0.40 \msun\ suggests a binary origin.  \citet{finley97} have failed to
see any flux excess up to 7500 \AA, suggesting that only a very cool
companion or a similar white dwarf could be present.  Our simulations with
both methods also reproduce the location of this object in Figure
\ref{fg:f10}, even though no radial velocity variations have been
detected by \citet{mmm2000}.  This object thus requires further
investigation.

{\it WD 0943$+$441 (G116-52)} --- This is a very interesting object.
First, both estimates of the ultraviolet temperature differ by more
than 5\% from the optical value (see Fig.~\ref{fg:f11}).  Second, its
location in Figure \ref{fg:f10} can be easily reproduced by our
simulations of unresolved binaries, suggesting that such temperature
discrepancies could be the result of a composite spectrum.  Finally,
the low mass of WD 0943$+$441 (0.39~\msun) strongly supports a binary
evolution for this object. Unfortunately, no radial velocity
measurements have been obtained for this star.  We thus suggest that
such measurements could reveal the binary nature of G116-52.
 
{\it WD 1022$+$050 (LP 550-52)} --- This is another known double
degenerate with measured radial velocity variations
\citep{mm1999}. This star shows
a temperature difference that exceeds $\sim10\%$ with both methods
used in our analysis (see Fig.~\ref{fg:f11}). Our simulations displayed in Figure
\ref{fg:f10} are able to reproduce the observed temperature differences 
for this star quite nicely, indicating that the UV and optical energy
distributions are altered by the presence of two stars in the
system. We thus confirm with our approach the binary nature of WD
1022$+$050.

{\it WD 1026$+$002 (PG)} --- This is a known binary consisting of a DA
star and a M-dwarf companion.  
Contrary to our optical spectrum, the spectrum of
\citet{saffer93} clearly shows the important contribution of the companion
beyond $\sim4500$ \AA, i.e. at wavelengths longer than those used here
to determine $\Topt$ and $\logg$.  Thus, the $V$ magnitude is most
likely contaminated by the companion, and this could easily explain the 
difference in temperatures reported here.

{\it WD 1031$-$114 (L825-14)} --- The difference between $\Tv$ and
$\Topt$ is a $\sim 2.3~\sigma$ result for this star. 
Furthermore, we can easily reproduce the measured temperature differences
with our simulations (Fig.~\ref{fg:f10}), 
and L825-14 thus represents a good double degenerate candidate,
despite the fact that no radial velocity
variations have been detected by \citet{mmm2000}.

{\it WD 2111$+$498 (GD 394)} --- GD 394 is a hot DA star ($39,000$ K)
that shows atmospheric abundance inhomogeneities and photometric
variations in the extreme UV portion of the spectrum.  From their UV
spectrum, \citet{dupuis2000} show that effective temperatures
determined from the UV continuum and from the Lyman line profiles
differ by $\sim4000$ K, and that these temperature differences are
likely due to dark spots at the surface of the star.  Given the facts
that no radial velocity variations have been observed by
\citet{saffer98} and that the line of sight is not significantly reddened,
no further conclusions can be reached for this star.

\section{CONCLUSION}\label{sect:concl}

In summary, we have gathered optical and UV spectra for 140 DA white
dwarfs for which we derived surface gravities and three different
effective temperature determinations. We then performed an internal
consistency check of all the effective temperature estimates and found
that the observed discrepancies were most likely due to interstellar
reddening. The presence of metals in the atmosphere of DA stars has
also been ruled out as a possible source of discrepancy since its
effect would have been in the opposite direction of what we
observed. Simulations of composite spectra have also shown that most
of our objects do not exhibit composite spectra, therefore ruling out
any binary nature.  In order to limit the effect of interstellar
reddening, we then restricted our sample to stars located within 75 pc
from the Sun, and which show statistically significant temperature
differences. This allowed us to identify interesting objects with
inconsistent optical and UV effective temperatures that deserve
further investigation.  In particular, the very low masses of WD
0509$-$007 (0.40 \msun) and WD 0943$+$441 (0.39 \msun) are almost
certainly indicative of past binary interactions and precise radial
velocity measurements should reveal the presence of a companion.  WD
1031$-$114 is also considered a good candidate on the basis of its
effective temperature differences and clearly deserves further
investigation.

We thank I.~Hubeny for providing us with his synthetic spectrum of
G191-B2B, and A.~Gianninas for a careful reading of our manuscript.
We also wish to thank the director and staff of the Steward
Observatory for the use of their facilities. This work was supported
in part by the NSERC Canada and by the Fund FQRNT
(Qu\'ebec). P. Bergeron is a Cottrell Scholar of Research Corporation.


\clearpage
\include{tab1}


\clearpage

\figcaption[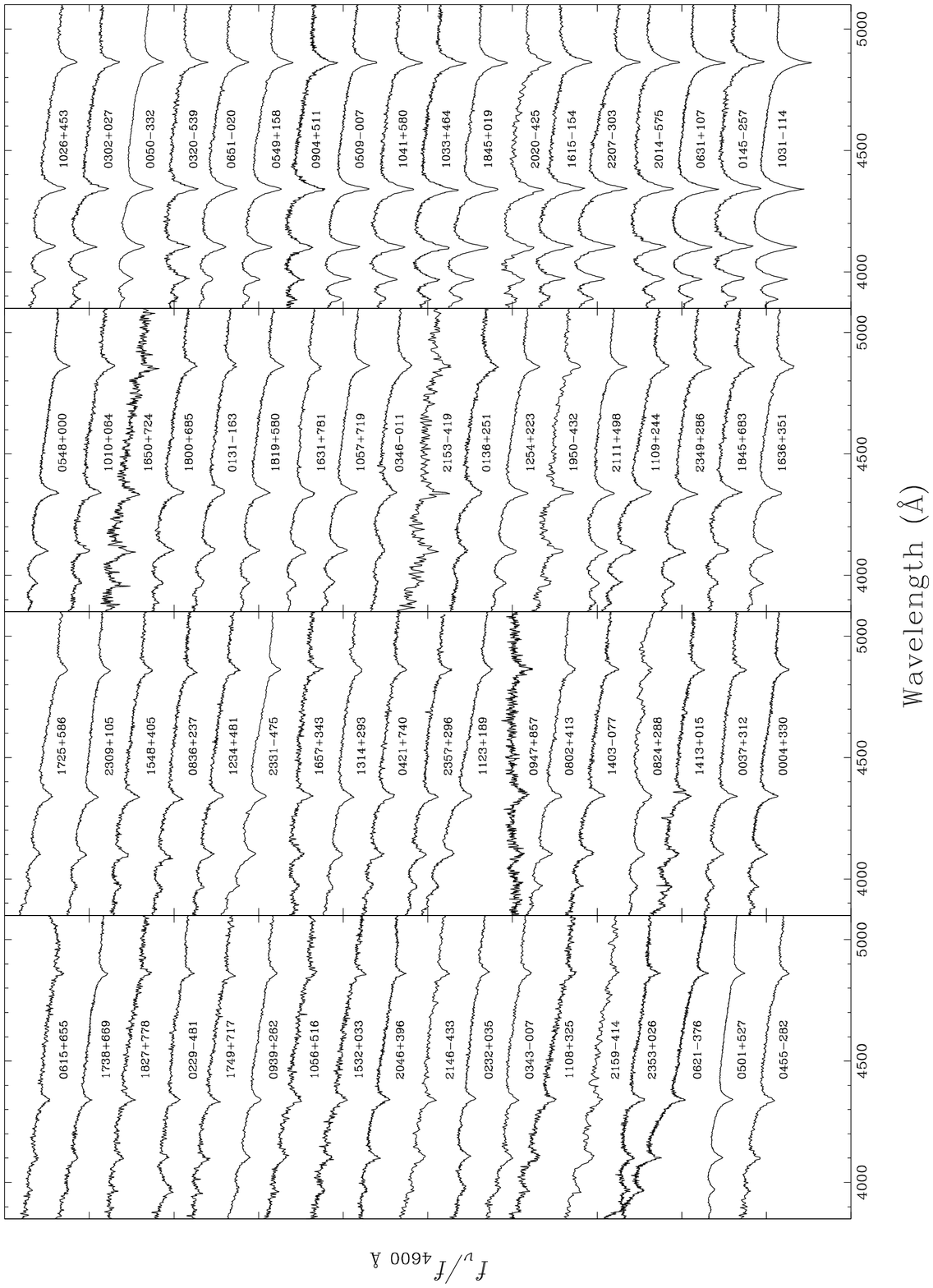] 
{Optical spectra for our sample of DA stars selected from the {\it
IUE} archive. The spectra are normalized at 4600 \AA\ and are shifted
vertically for clarity. The effective temperature decreases from upper
left to bottom right.\label{fg:f1}}

\figcaption[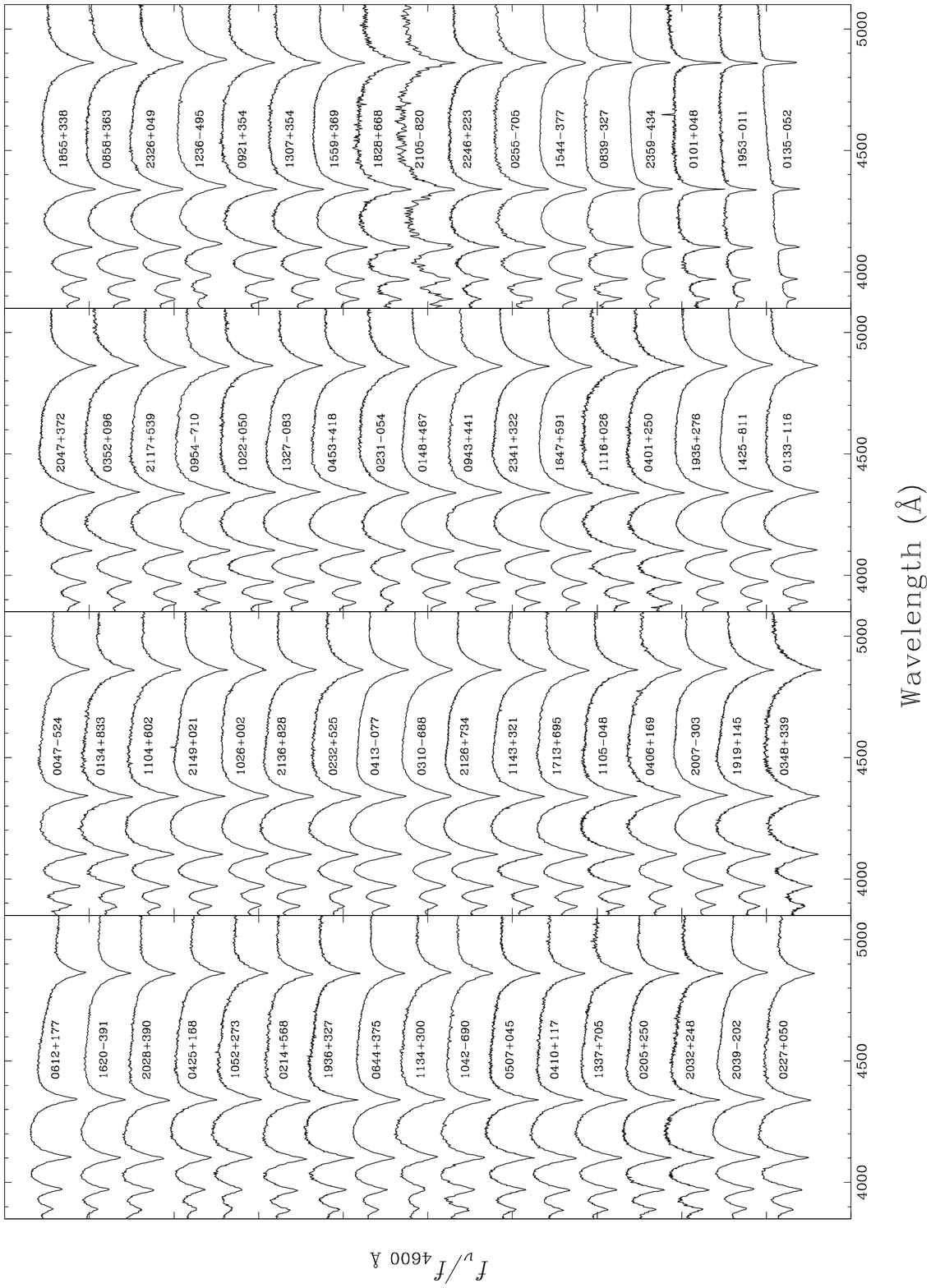] 
{Same as Fig.~\ref{fg:f1}.\label{fg:f2}} 

\figcaption[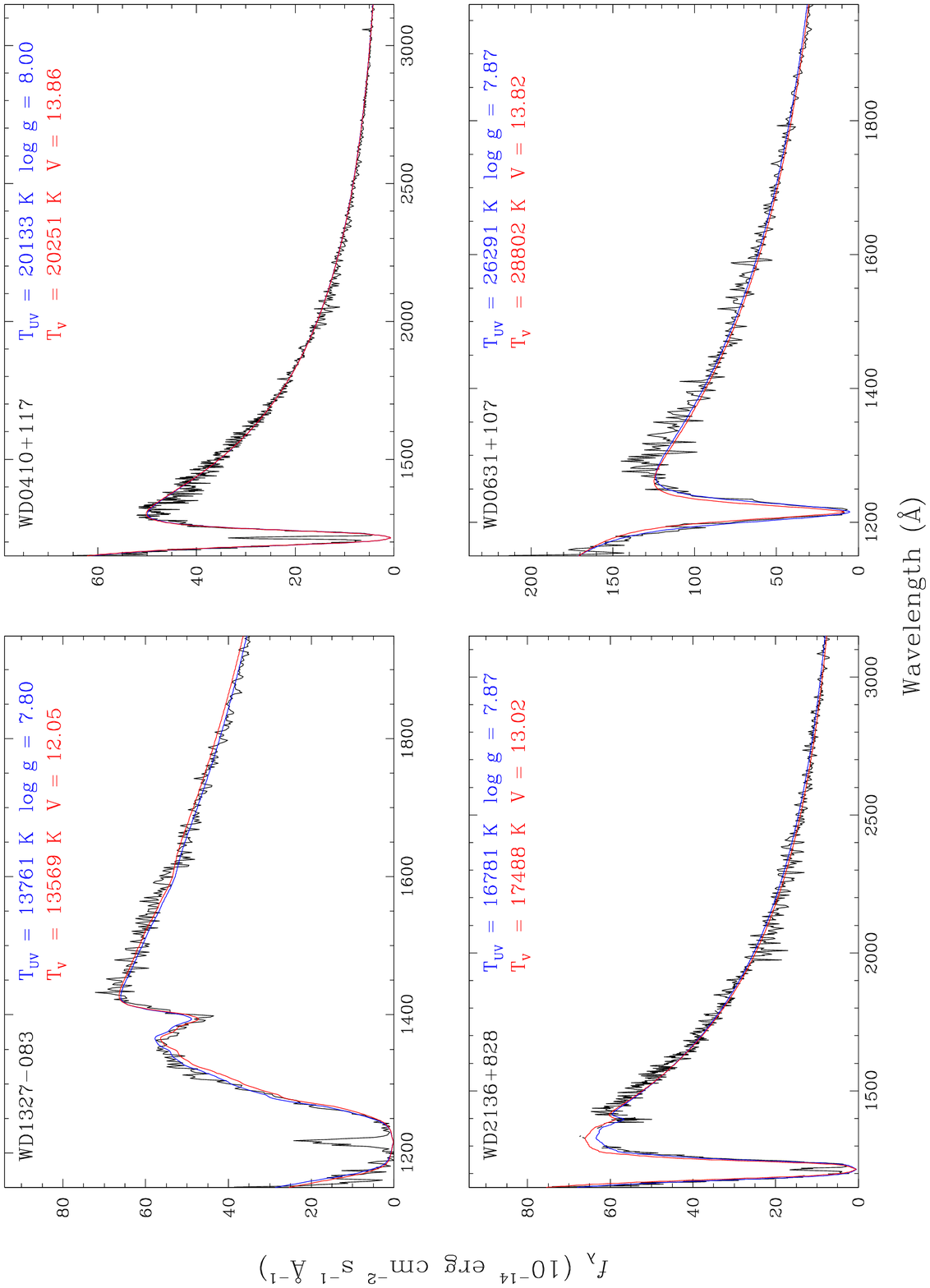] 
{Sample fits to the UV energy distribution using the slope method
({\it blue}) and the $V$-normalization method ({\it red}) to determine $\Te$. In
both cases, we assume the $\logg$ value obtained from the optical
spectrum.\label{fg:f3}}

\figcaption[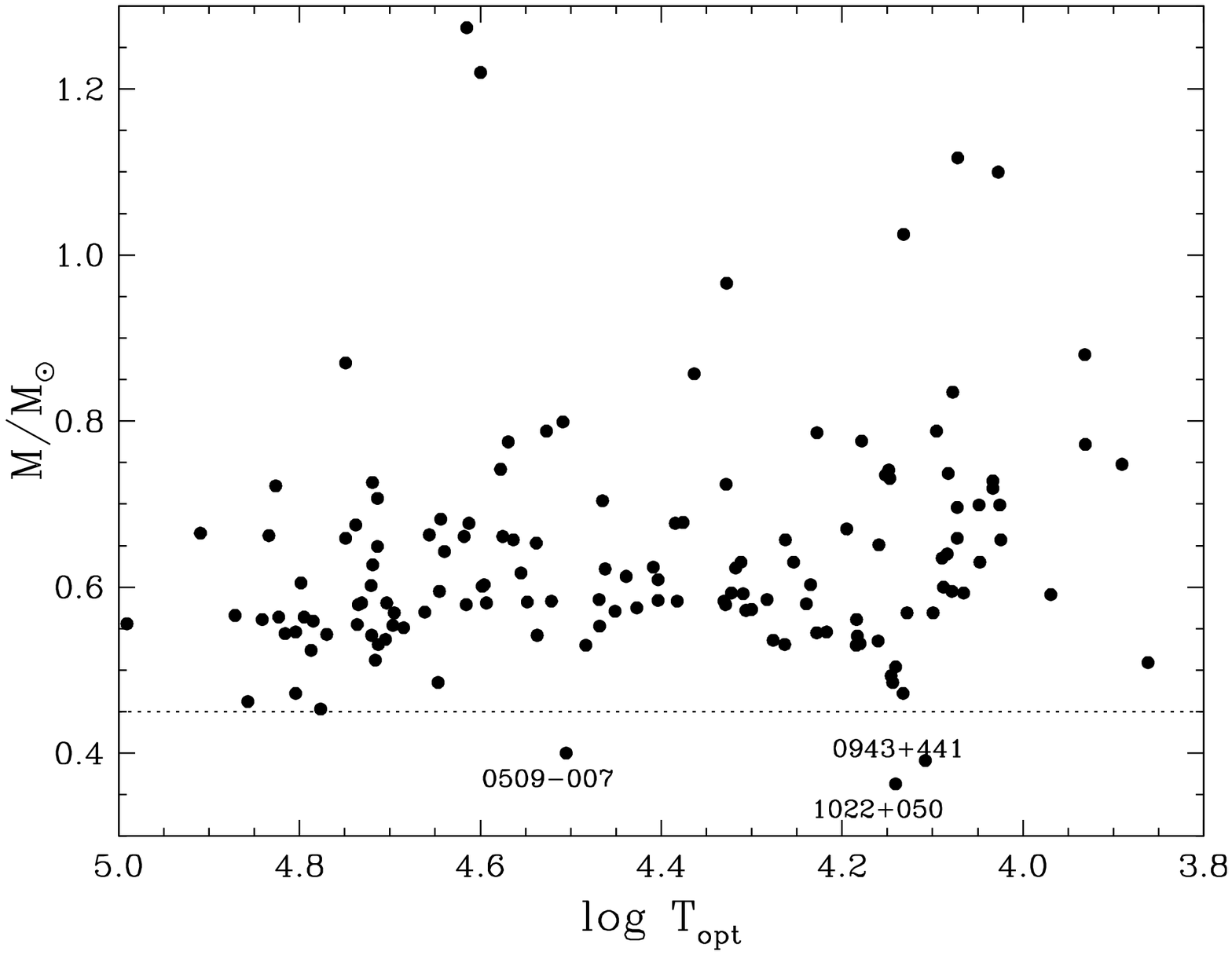] 
{Mass distribution for all the DA stars in our sample as a function
of $T_{\rm opt}$.  The objects labeled in the
figure correspond to low-mass stars ($M\ \lta 0.45$~\msun), which must
have evolved in a binary system since the main-sequence lifetime of
their progenitor is estimated to be longer than the age of the Galaxy.
\label{fg:f4}}

\figcaption[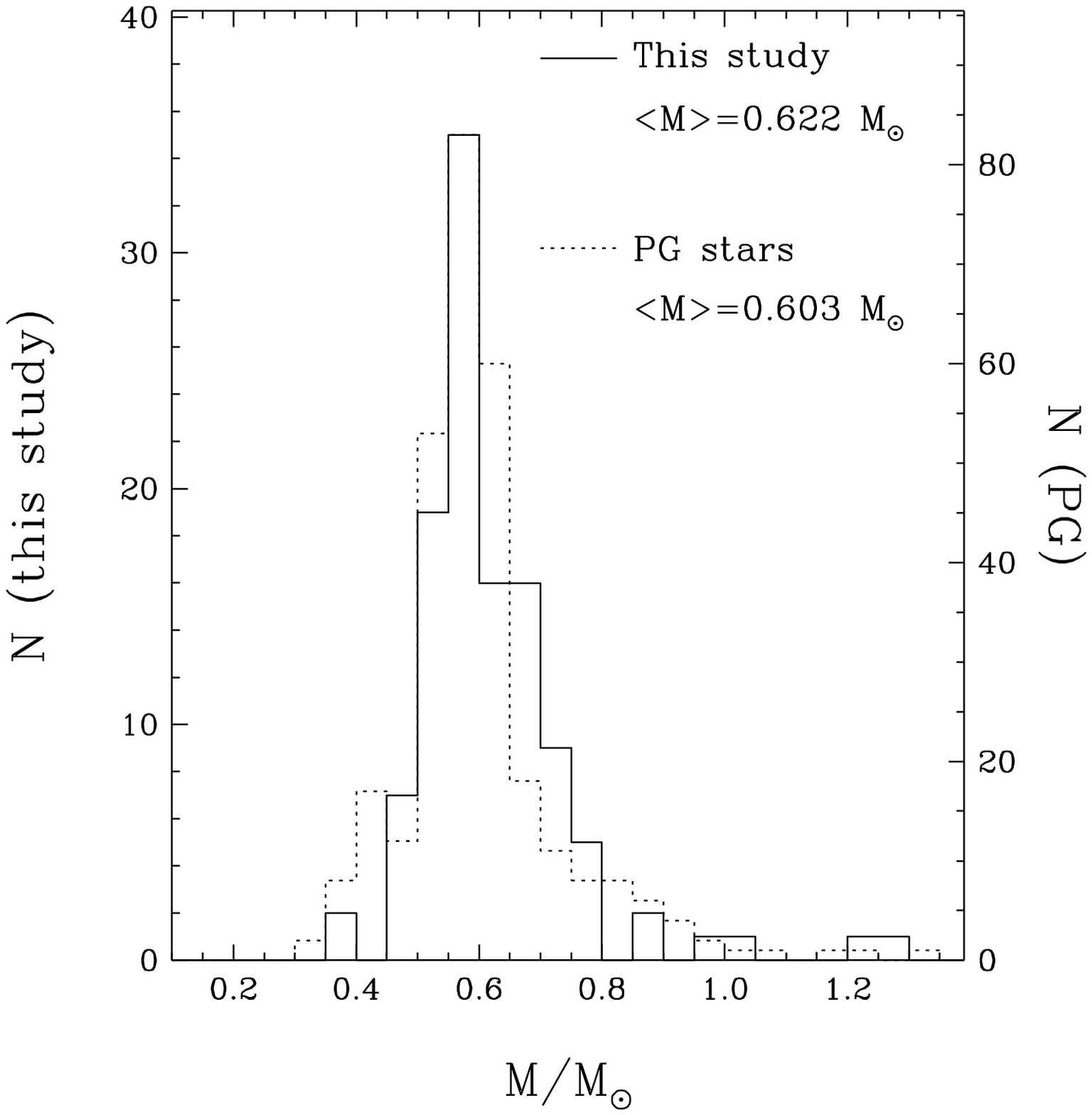] 
{Mass distribution for the DA stars in our sample hotter than 13,000 K
compared to the mass distribution obtained by LBH05 for the DA stars
in the PG survey in the same temperature range.  \label{fg:f5}}

\figcaption[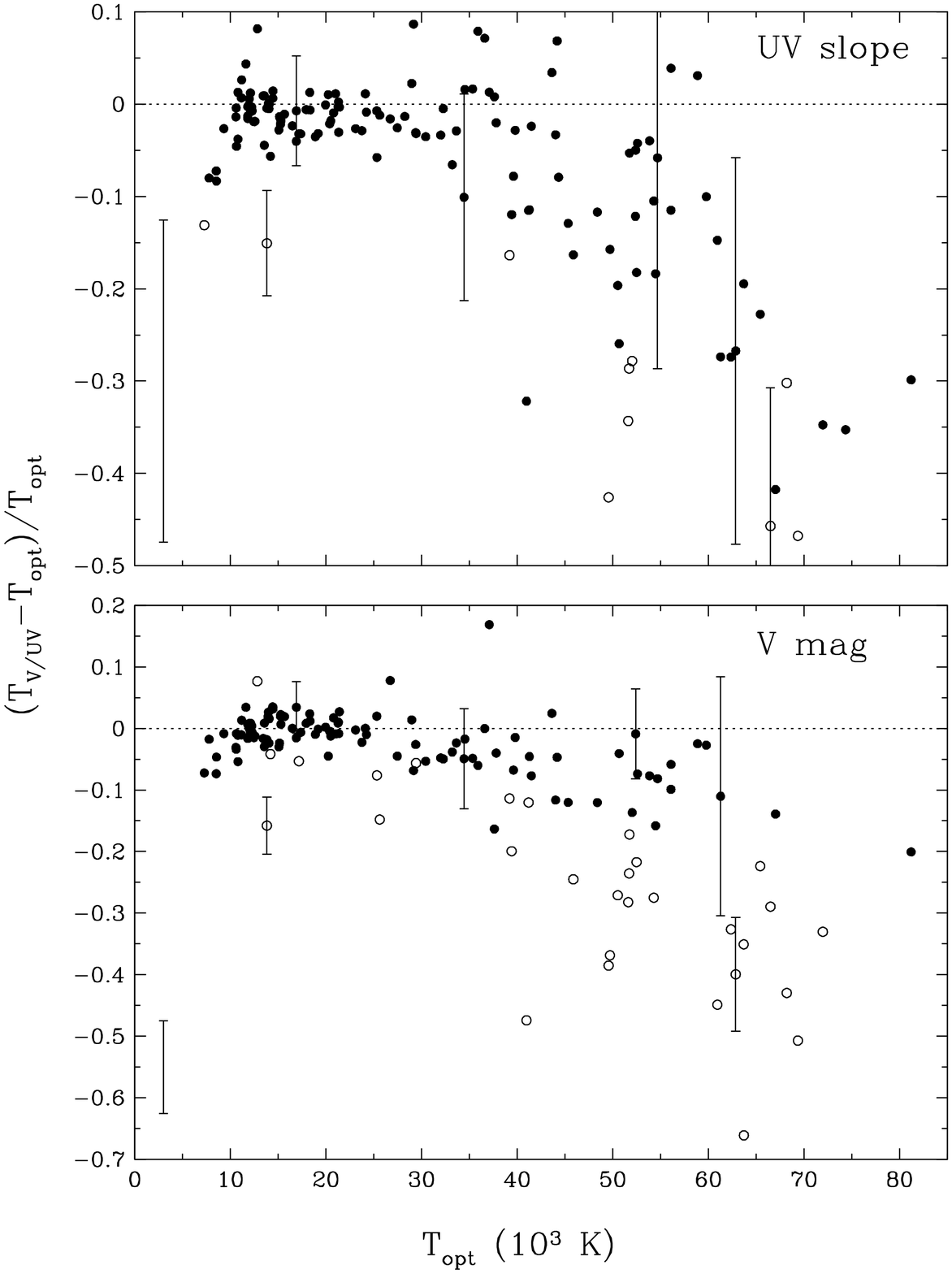] 
{Differences in optical $(\Topt)$ and ultraviolet ($T_{V/{\rm UV}}$)
effective temperatures (normalized to $\Topt$) as a function of
$\Topt$. The UV-slope and $V$-normalization methods are used in the
top and bottom panel, respectively. Open circles represent objects for
which optical and UV temperature estimates are inconsistent within the
uncertainties. A few indicative error bars are also shown, together
with the average error bar in the lower left corner of each panel.
\label{fg:f6}}

\figcaption[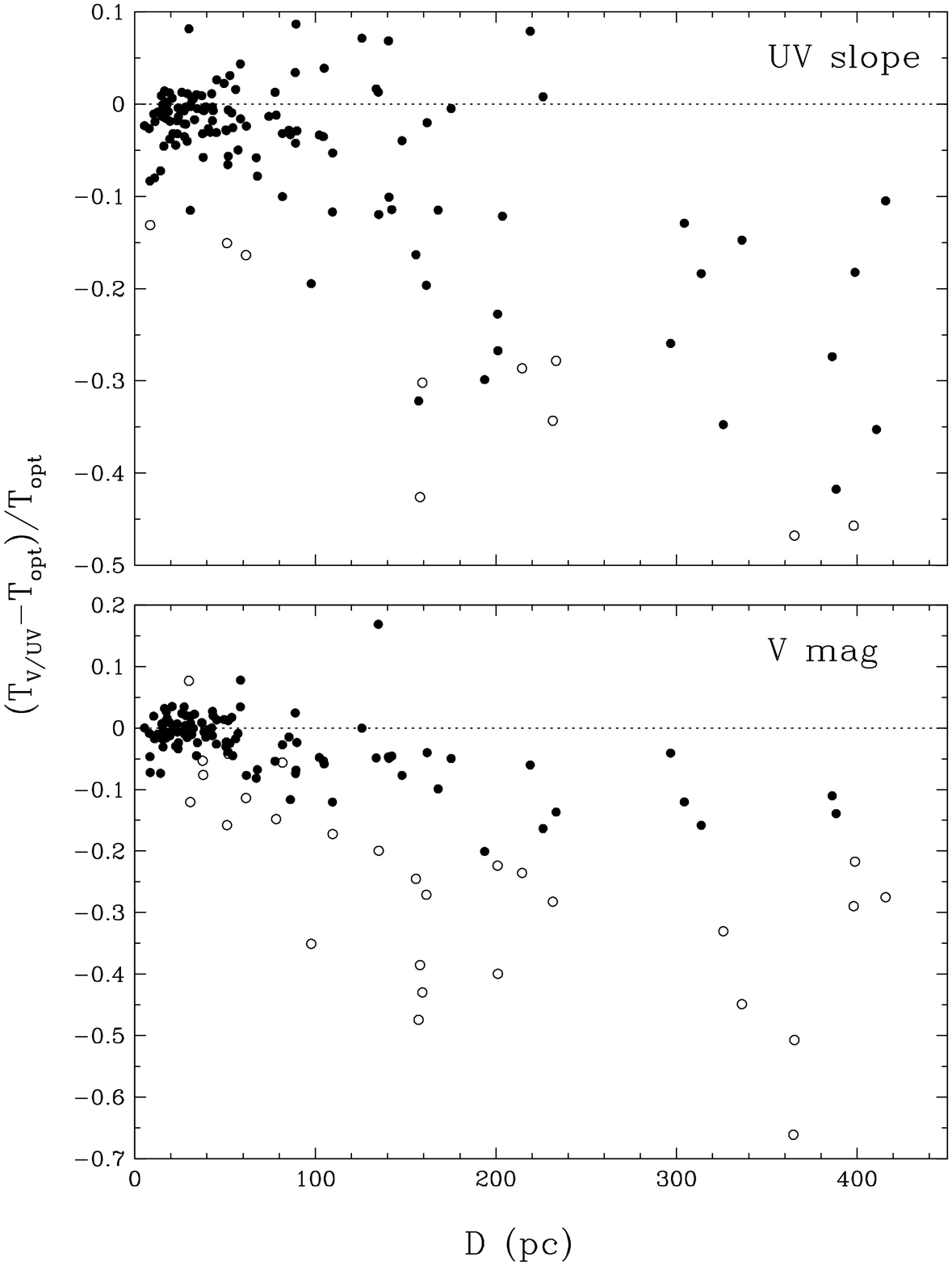] 
{Same as Fig.~\ref{fg:f6} but as a function of distance.\label{fg:f7}}

\figcaption[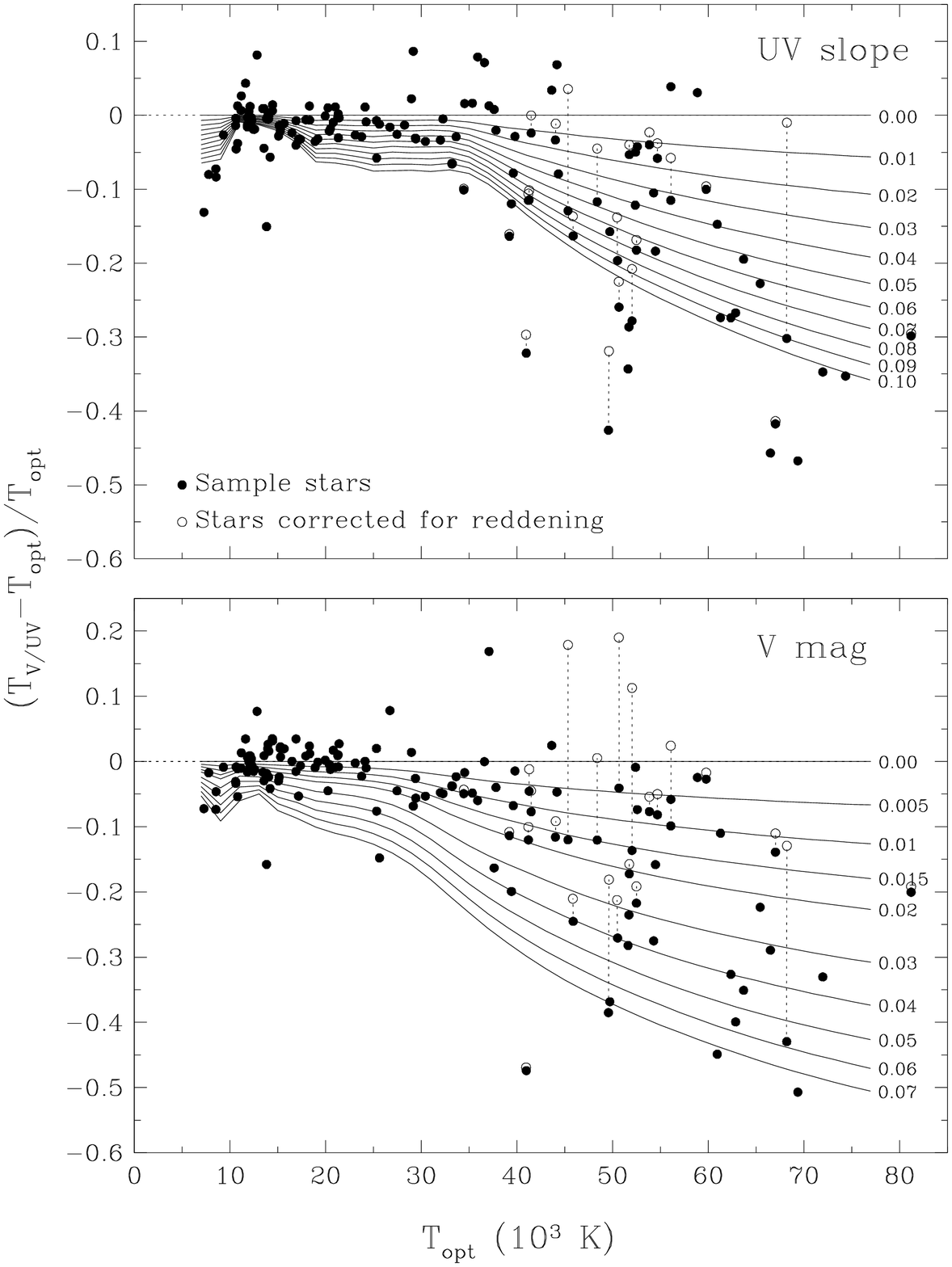] 
{Effect of interstellar reddening on the temperature differences as
determined from the UV-slope ({\it top}) and $V$-normalization ({\it
bottom}) methods.  The color excess $E(B-V)$ for each simulation is
indicated on the right side of each curve.  Results for our DA sample
are shown by filled circles, while objects corrected for reddening are
represented by open circles and connected by dashed lines to the
corresponding unreddened solutions.  \label{fg:f8}}

\figcaption[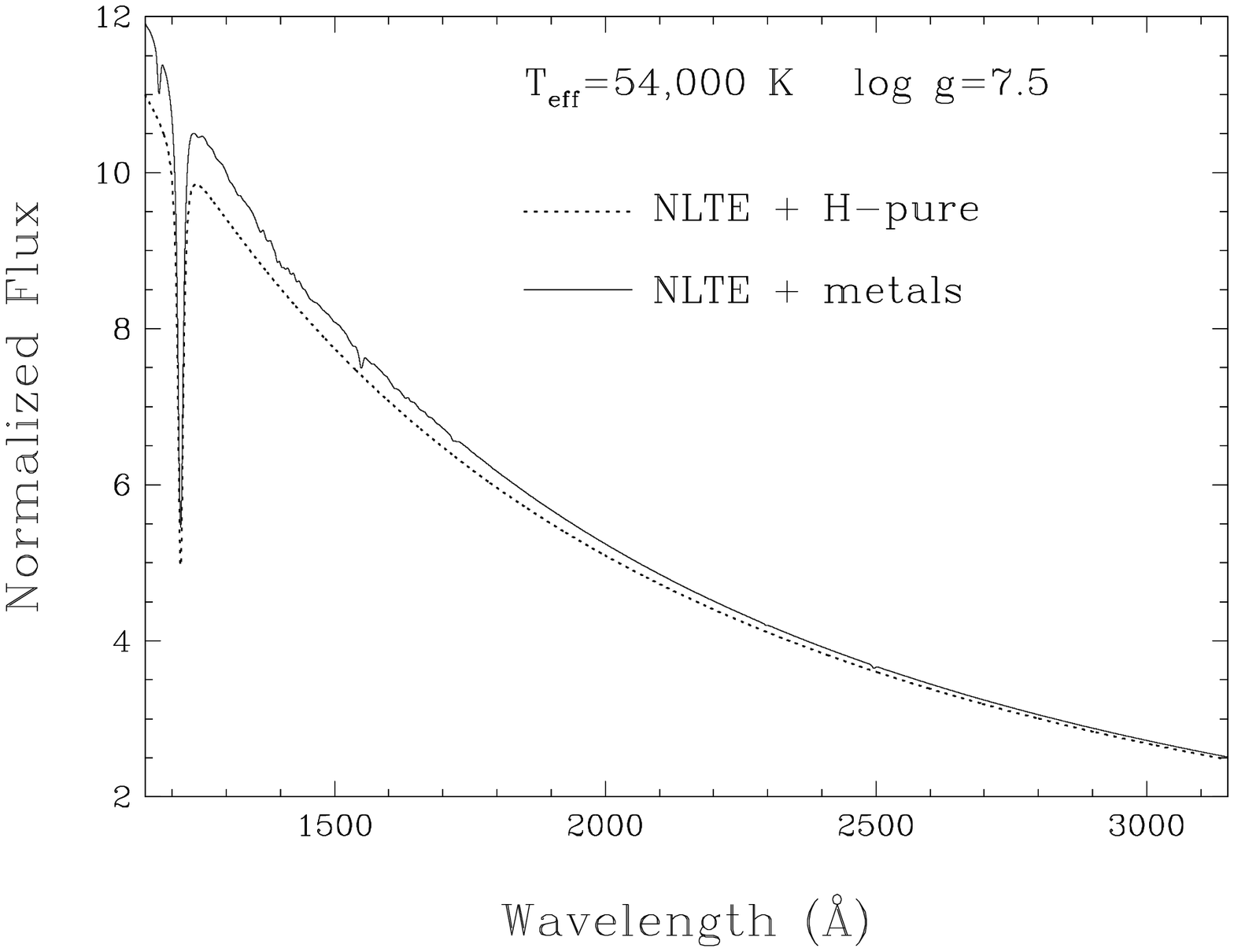] 
{Synthetic spectrum of a model at $\Te=54,000$~K and $\logg=7.5$ that
includes H, He, C, N, O, Si, Fe, and Ni ({\it solid line}) compared to
a pure hydrogen synthetic spectrum with the same atmospheric
parameters ({\it dotted line}).  Both spectra have been normalized at
5500 \AA\ and convolved with a 6 \AA\ FWHM instrumental
profile. \label{fg:f9}}

\figcaption[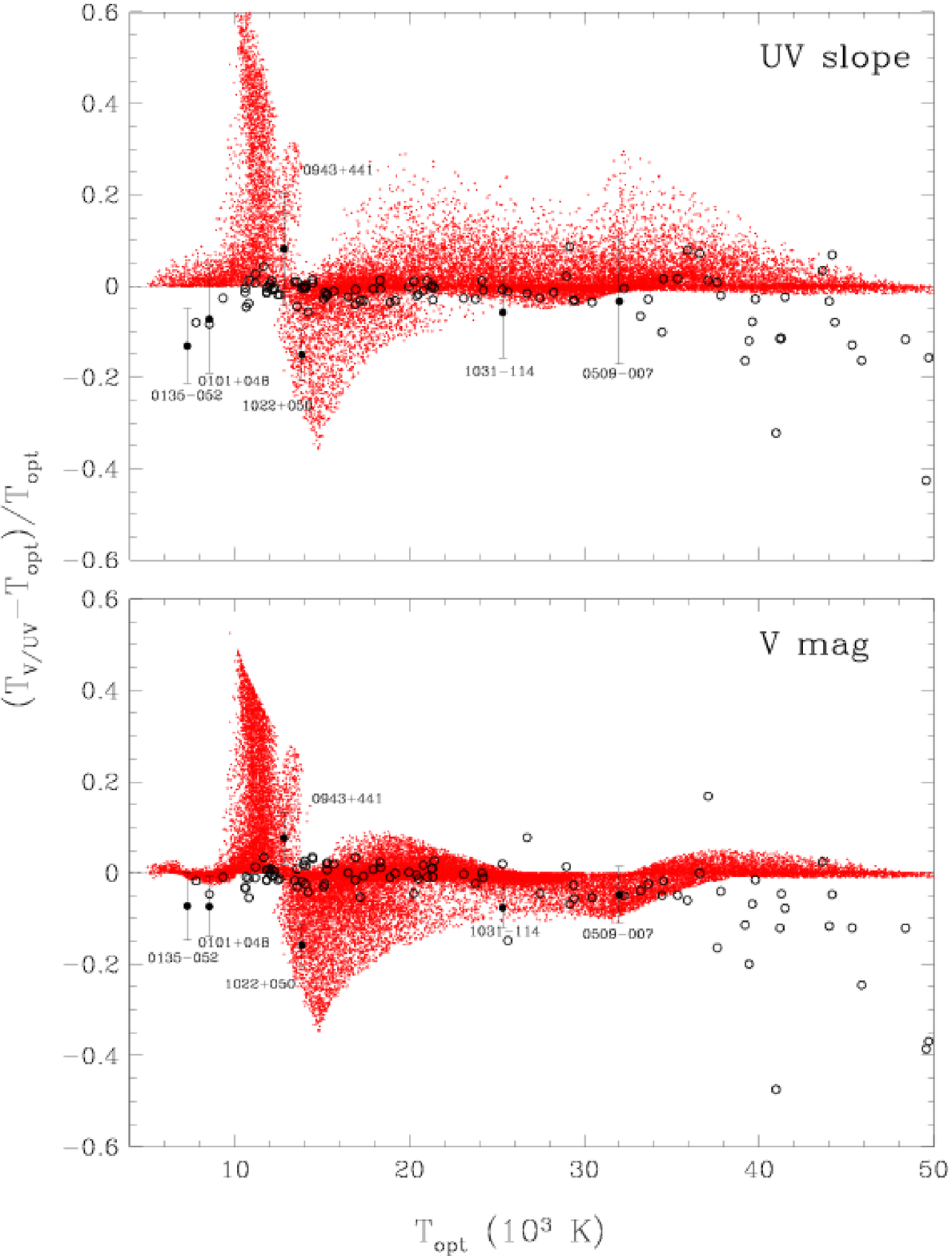] 
{Simulated temperature differences for unresolved binary white dwarf
systems using the UV-slope method ({\it top}) and the
$V$-normalization method ({\it bottom}).  Small red dots correspond to
our simulations while the results for our sample of DA stars are
represented by circles. Filled circles with labels and error bars are
discussed in \S~\ref{sect:objects}. [{\it See the electronic version of
the Journal for a color version of this figure.}]\label{fg:f10}}

\figcaption[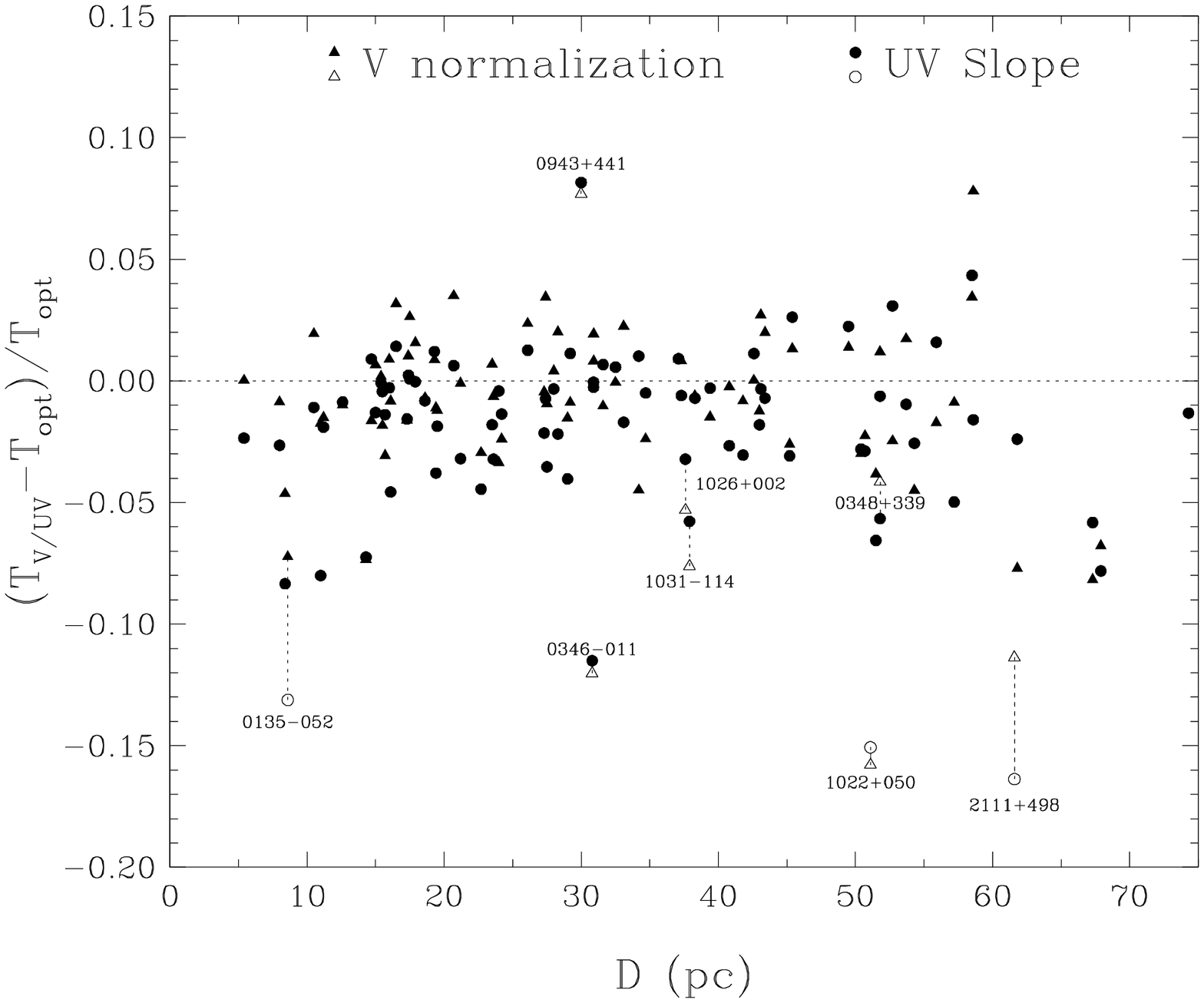] 
{Differences in optical and UV temperatures for stars located within
75 pc. Both the UV-slope method ({\it circles}) and the
$V$-normalization method ({\it triangles}) are shown here. Objects
with inconsistent temperatures are shown with open symbols,
labeled by their WD number, and discussed in \S~\ref{sect:objects}.
\label{fg:f11}}

\clearpage
\begin{figure}[p]
  \plotone{f1.eps}
  \begin{flushright}
    Figure \ref{fg:f1}
  \end{flushright}
\end{figure}

\begin{figure}[p]
  \plotone{f2.eps}
  \begin{flushright}
    Figure \ref{fg:f2}
  \end{flushright}
\end{figure}

\begin{figure}[p]
  \plotone{f3.eps}
  \begin{flushright}
    Figure \ref{fg:f3}
  \end{flushright}
\end{figure}

\begin{figure}[p]
  \plotone{f4.eps}
  \begin{flushright}
    Figure \ref{fg:f4}
  \end{flushright}
\end{figure}

\begin{figure}[p]
  \plotone{f5.eps}
  \begin{flushright}
    Figure \ref{fg:f5}
  \end{flushright}
\end{figure}

\begin{figure}[p]
  \plotone{f6.eps}
  \begin{flushright}
    Figure \ref{fg:f6}
  \end{flushright}
\end{figure}

\begin{figure}[p]
  \plotone{f7.eps}
  \begin{flushright}
    Figure \ref{fg:f7}
  \end{flushright}
\end{figure}

\begin{figure}[p]
  \plotone{f8.eps}
  \begin{flushright}
    Figure \ref{fg:f8}
  \end{flushright}
\end{figure}

\begin{figure}[p]
  \plotone{f9.eps}
  \begin{flushright}
    Figure \ref{fg:f9}
  \end{flushright}
\end{figure}

\begin{figure}[p]
  \plotone{f10.eps}
  \begin{flushright}
    Figure \ref{fg:f10}
  \end{flushright}
\end{figure}

\begin{figure}[p]
  \plotone{f11.eps}
  \begin{flushright}
    Figure \ref{fg:f11}
  \end{flushright}
\end{figure}

\end{document}

%% file: tab1.tex
\begin{deluxetable}{llccc r@{}c@{}l r@{}c@{}l cc}
\tabletypesize{\scriptsize}
\tablecolumns{13}
\tablewidth{0pt}
\tablecaption{Atmospheric Parameters for IUE DA White Dwarf Stars}
\tablehead{
\colhead{} &
\colhead{} &
\colhead{} &
\colhead{$T_{\textrm{opt}}$} &
\colhead{} &
\colhead{} &
\colhead{$T_{\textrm{UV}}$} &
\colhead{} &
\colhead{} &
\colhead{$T_V$} &
\colhead{} &
\colhead{} &
\colhead{$D$} \\
\colhead{WD} &
\colhead{Name} &
\colhead{$V$} &
\colhead{(K)} &
\colhead{log $\textit{g}$} &
\colhead{} &
\colhead{(K)} &
\colhead{} &
\colhead{} &
\colhead{(K)} &
\colhead{} &
\colhead{$M$/$M_{\odot}$} &
\colhead{(pc)} 
}
\startdata
0004$+$330& GD 2           &13.82&  48385& 7.68&  42726& $\pm$& 9600&  42555& $\pm$& 3900& 0.55&  109.5\\
0037$+$312& GD 8           &14.66&  49564& 7.72&  28452& $\pm$& 5400&  30474& $\pm$& 3500& 0.57&  158.0\\
0047$-$524& BPM 16274      &14.20&  18339& 7.83&  18223& $\pm$&  900&  18559& $\pm$&  350& 0.53&   51.8\\
0050$-$332& GD 659         &13.36&  34529& 7.99&  35077& $\pm$& 4000&  33936& $\pm$& 2000& 0.65&   55.9\\
0101$+$048& G2-17        &13.96&   8530& 8.27&   7912& $\pm$&  900&   7903& $\pm$&  450& 0.77&   14.3\\
0131$-$163& GD 984         &13.98&  44015& 8.00&  42553& $\pm$& 8400&  38905& $\pm$& 3700& 0.68&   86.2\\
0133$-$116& Ross 548       &14.16&  11986& 7.98&  12054& $\pm$&  800&  11979& $\pm$&  400& 0.60&   32.5\\
0134$+$833& GD 419         &13.06&  18311& 8.06&  18542& $\pm$& 1000&  18744& $\pm$&  400& 0.66&   26.1\\
0135$-$052& L870-2       &12.84&   7273& 7.85&   6319& $\pm$&  500&   6747& $\pm$&  450& 0.51&    8.6\\
0136$+$251& PG 0136$+$251  &16.00&  39791& 9.03&  38660& $\pm$& 5000&  39209& $\pm$& 6200& 1.22&   85.4\\
0145$-$257& GD 1401        &14.69&  25625& 7.97&  25319& $\pm$& 2600&  21836& $\pm$&  975& 0.62&   78.3\\
0148$+$467& GD 279         &12.17&  13432& 7.93&  13553& $\pm$&  400&  13213& $\pm$&  200& 0.57&   14.7\\
0205$+$250& G35-29       &13.23&  20243& 7.90&  20449& $\pm$& 2500&  19334& $\pm$&  600& 0.57&   34.2\\
0214$+$568& H Per 1166     &13.65&  21408& 7.91&  21338& $\pm$& 1500&  21990& $\pm$&  500& 0.58&   43.1\\
0227$+$050& Feige 22       &12.79&  18887& 7.84&  18219& $\pm$& 1250&  18708& $\pm$&  450& 0.54&   27.5\\
0229$-$481& LB 1628        &14.53&  71970& 7.09&  46972& $\pm$&17500&  48200& $\pm$& 8250& 0.46&  325.9\\
0231$-$054& GD 31          &14.24&  13552& 8.66&  12949& $\pm$&  400&  13152& $\pm$&  200& 1.02&   22.7\\
0232$+$035& Feige 24       &12.40&  63698& 7.25&  51306& $\pm$&14500&  41356& $\pm$& 2250& 0.47&   97.7\\
0232$+$525& G174-5       &13.75&  16892& 8.27&  16211& $\pm$& 1000&  16633& $\pm$&  425& 0.79&   29.0\\
0255$-$705& BPM 2819       &14.08&  10608& 8.15&  10564& $\pm$& 2500&  10251& $\pm$& 1275& 0.70&   24.0\\
0302$+$027& Feige 31       &14.97&  35337& 7.84&  35916& $\pm$&10000&  33620& $\pm$& 2800& 0.58&  133.7\\
0310$-$688& LB 3303        &11.40&  15658& 8.09&  15487& $\pm$&  600&  15962& $\pm$&  325& 0.67&   10.5\\
0320$-$539& LB 1663        &14.99&  34443& 7.75&  30966& $\pm$& 3400&  32745& $\pm$& 2375& 0.54&  140.8\\
0343$-$007& KUV 03439$-$0048  &14.91&  62859& 7.73&  46058& $\pm$&12200&  37759& $\pm$& 4750& 0.61&  201.1\\
0346$-$011& GD 50          &13.99&  41196& 9.15&  36456& $\pm$& 3800&  36241& $\pm$& 1550& 1.27&   30.8\\
0348$+$339& GD 52          &15.20&  14194& 8.20&  13391& $\pm$&  600&  13605& $\pm$&  300& 0.74&   51.8\\
0352$+$096& HZ 4           &14.34&  14033& 8.19&  13964& $\pm$&  600&  13699& $\pm$&  250& 0.73&   34.7\\
0401$+$250& G8-8         &13.81&  12240& 7.99&  12200& $\pm$&  400&  12291& $\pm$&  250& 0.60&   28.0\\
0406$+$169& LB 227         &15.13&  15073& 8.26&  14650& $\pm$&  600&  14624& $\pm$&  300& 0.78&   50.4\\
0410$+$117& HZ 2           &13.86&  20504& 8.01&  20134& $\pm$& 1200&  20251& $\pm$&  400& 0.63&   43.0\\
0413$-$077& 40 Eri B       & 9.52&  16480& 7.87&  16093& $\pm$&  600&  16483& $\pm$&  300& 0.55&    5.4\\
0421$+$740& RE J0427$+$741 & --- &  52372& 7.85&  46006& $\pm$&16000&       &  --- &     & 0.63&  203.6\\
0425$+$168& GH 7-233     &14.06&  23760& 8.08&  23076& $\pm$& 2200&  23224& $\pm$&  950& 0.68&   50.7\\
0453$+$418& GD 64          &13.89&  13564& 7.74&  13688& $\pm$&  700&  13684& $\pm$&  400& 0.47&   37.1\\
0455$-$282& RE J0457$-$280 &13.95&  56087& 7.90&  58265& $\pm$&20000&  52821& $\pm$& 5450& 0.66&  104.9\\
0501$+$527& G191-B2B     &11.78&  58865& 7.57&  60680& $\pm$&15000&  57414& $\pm$& 4700& 0.54&   52.7\\
0507$+$045& HS 0507$+$0435A &14.30&  20787& 7.99&  20587& $\pm$& 2800&  21147& $\pm$&  700& 0.62&   53.7\\
0509$-$007& RE J0512$-$004 &13.80&  32004& 7.37&  30931& $\pm$& 4000&  30475& $\pm$& 1575& 0.40&  102.2\\
0548$+$000& GD 257         &14.77&  45871& 7.75&  38384& $\pm$& 8800&  34631& $\pm$& 3600& 0.57&  155.7\\
0549$+$158& GD 71          &13.04&  33212& 7.85&  31034& $\pm$& 2200&  31943& $\pm$& 1000& 0.58&   51.5\\
0612$+$177& G104-27      &13.39&  25312& 7.94&  25132& $\pm$& 1800&  25817& $\pm$&  675& 0.61&   43.4\\
0615$+$655& HS 0615$+$6535 &15.70&  97889& 7.12&  63424& $\pm$&39500&  27217& $\pm$& 4550& 0.56&  682.3\\
0621$-$376& RE J0623$-$374 &12.09&  59779& 7.24&  53791& $\pm$&20500&  58154& $\pm$&11625& 0.45&   81.8\\
0631$+$107& WD 0631$+$107  &13.82&  26718& 7.87&  26291& $\pm$& 2600&  28803& $\pm$& 1650& 0.57&   58.6\\
0644$+$375& G87-7        &12.07&  21300& 8.16&  21349& $\pm$& 2400&  21519& $\pm$&  450& 0.72&   17.4\\
0651$-$020& GD 80          &14.83&  33643& 8.23&  32669& $\pm$& 3400&  32853& $\pm$& 1775& 0.79&   89.8\\
0802$+$413& KUV 08026+4118 &15.21&  51616& 7.59&  33905& $\pm$& 6400&  37049& $\pm$& 4400& 0.53&  231.5\\
0824$+$288& PG 0824$+$289  &14.73&  50525& 7.75&  40603& $\pm$& 7000&  36841& $\pm$& 2275& 0.58&  161.5\\
0836$+$237& PG 0836$+$237  &16.64&  54290& 7.71&  48595& $\pm$&27500&  39357& $\pm$& 4825& 0.58&  415.8\\
0839$-$327& LHS 253        &11.90&   9318& 7.99&   9071& $\pm$&  400&   9238& $\pm$&  300& 0.59&    8.0\\
0858$+$363& GD 99          &14.76&  11825& 8.09&  11790& $\pm$&  650&  11650& $\pm$&  400& 0.66&   39.4\\
0904$+$511& PG 0904$+$512  &16.40&  32268& 8.25&  32111& $\pm$& 9500&  30669& $\pm$& 2650& 0.80&  175.2\\
0921$+$354& G117-B15A    &15.50&  11627& 7.98&  12132& $\pm$&  800&  12028& $\pm$&  350& 0.59&   58.5\\
0939$+$262& PG 0939$+$262  &14.53&  68201& 7.84&  47613& $\pm$&12400&  38903& $\pm$& 2950& 0.66&  159.3\\
0943$+$441& G116-52      &13.29&  12822& 7.55&  13868& $\pm$& 1400&  13807& $\pm$&  550& 0.39&   30.0\\
0947$+$857& RE J0957$+$852 &15.80&  51709& 8.02&  36908& $\pm$& 8000&  39539& $\pm$& 4125& 0.71&  214.5\\
0954$-$710& BPM 6082       &13.48&  13930& 7.76&  13894& $\pm$&  600&  14200& $\pm$&  425& 0.49&   30.9\\
1010$+$064& PG 1010$+$065  &16.61&  45329& 7.96&  39478& $\pm$&17000&  39885& $\pm$& 7575& 0.66&  304.3\\
1022$+$050& LP 550-52    &14.18&  13828& 7.47&  11745& $\pm$&  600&  11645& $\pm$&  450& 0.36&   51.1\\
1026$+$002& PG 1026$+$002  &13.83&  17172& 7.97&  16620& $\pm$& 1100&  16261& $\pm$&  400& 0.60&   37.6\\
1026$+$453& PG 1026$+$454  &16.13&  35900& 7.91&  38734& $\pm$&12000&  33748& $\pm$& 3300& 0.62&  219.0\\
1031$-$114& L825-14      &13.02&  25328& 7.89&  23865& $\pm$& 2200&  23399& $\pm$&  800& 0.58&   37.9\\
1033$+$464& GD 123         &14.34&  29425& 7.88&  28487& $\pm$& 2600&  27772& $\pm$&  850& 0.58&   81.8\\
1041$+$580& PG 1041$+$580  &14.60&  30436& 7.75&  29363& $\pm$& 4800&  28813& $\pm$& 1525& 0.53&  104.4\\
1042$-$690& BPM 6502       &12.87&  21012& 7.93&  21250& $\pm$& 1800&  20826& $\pm$&  600& 0.59&   29.2\\
1052$+$273& GD 125         &14.11&  23095& 8.37&  22480& $\pm$& 1200&  23039& $\pm$&  450& 0.86&   40.8\\
1056$+$516& LB 1919        &16.76&  67022& 7.99&  39045& $\pm$&18000&  57701& $\pm$&18400& 0.72&  388.4\\
1057$+$719& PG 1057$+$719  &14.80&  41276& 7.80&  36555& $\pm$& 6200&  39394& $\pm$& 3600& 0.58&  142.3\\
1104$+$602& WD 1104$+$602  &13.80&  17922& 8.02&  17815& $\pm$& 1600&  18070& $\pm$&  575& 0.63&   37.3\\
1105$-$048& L970-30      &12.92&  15142& 7.85&  14936& $\pm$&  600&  14780& $\pm$&  300& 0.53&   24.2\\
1108$+$325& PG 1108$+$325  &16.80&  62364& 7.61&  45281& $\pm$&21500&  42007& $\pm$& 6500& 0.56&  530.5\\
1109$+$244& PG 1109$+$244  &15.77&  37812& 8.14&  37047& $\pm$&10800&  36305& $\pm$& 4175& 0.74&  161.9\\
1116$+$026& PG 1116$+$026  &14.57&  12286& 8.05&  12198& $\pm$&  300&  12211& $\pm$&  150& 0.63&   38.3\\
1123$+$189& PG 1123$+$189  &14.13&  51751& 7.90&  49004& $\pm$&12800&  42837& $\pm$& 3575& 0.65&  109.6\\
1134$+$300& GD 140         &12.52&  21259& 8.55&  21199& $\pm$&  800&  21447& $\pm$&  400& 0.97&   16.0\\
1143$+$321& G148-7       &13.66&  15276& 7.90&  15016& $\pm$&  650&  15619& $\pm$&  425& 0.56&   33.1\\
1234$+$481& HS 1234$+$4811 &14.42&  53843& 7.72&  51701& $\pm$&14200&  49703& $\pm$& 5700& 0.58&  148.0\\
1236$-$495& BPM 37093      &13.96&  11809& 8.84&  11656& $\pm$&  650&  11887& $\pm$&  425& 1.12&   15.0\\
1254$+$223& GD 153         &13.35&  39615& 7.86&  36519& $\pm$& 4200&  36932& $\pm$& 2100& 0.60&   67.9\\
1307$+$354& GD 154         &15.31&  11180& 8.15&  11473& $\pm$&  800&  11328& $\pm$&  500& 0.70&   45.4\\
1314$+$293& HZ 43          &12.98&  52394& 8.06&  49785& $\pm$&10600&  51931& $\pm$& 3200& 0.73&   57.2\\
1327$-$083& Wolf 485       &12.05&  13823& 7.80&  13762& $\pm$&  300&  13570& $\pm$&  150& 0.50&   15.5\\
1337$+$705& G238-44      &12.78&  20390& 7.94&  19954& $\pm$& 1000&  20298& $\pm$&  300& 0.59&   27.3\\
1403$-$077& PG 1403$-$077  &15.82&  50664& 7.62&  37519& $\pm$&17000&  48595& $\pm$&10300& 0.54&  296.7\\
1413$+$015& PG 1413$+$015  &17.01&  49716& 7.68&  41897& $\pm$&14500&  31394& $\pm$& 3350& 0.55&  483.1\\
1425$-$811& L19-2        &13.35&  12098& 8.21&  12244& $\pm$&  350&  12203& $\pm$&  200& 0.74&   19.3\\
1532$+$033& PG 1532$+$033  &16.02&  66495& 7.57&  36114& $\pm$& 8800&  47257& $\pm$& 7700& 0.56&  398.1\\
1544$-$377& L481-60      &13.07&  10583& 8.09&  10436& $\pm$&  400&  10259& $\pm$&  275& 0.66&   15.7\\
1548$+$405& PG 1548$+$405  &15.89&  54476& 7.64&  44464& $\pm$&19500&  45863& $\pm$& 7200& 0.56&  313.7\\
1559$+$369& G180-23      &14.36&  11160& 8.04&  11235& $\pm$&  750&  11045& $\pm$&  525& 0.63&   31.6\\
1615$-$154& G153-41      &13.40&  28971& 7.95&  29620& $\pm$& 2800&  29373& $\pm$& 1125& 0.62&   49.5\\
1620$-$391& CD $-$38 10980 &10.99&  24231& 8.07&  24020& $\pm$& 1000&  23991& $\pm$&  400& 0.68&   12.6\\
1631$+$781& WD 1631$+$781  &13.28&  41489& 7.97&  40496& $\pm$& 7000&  38293& $\pm$& 2800& 0.66&   61.8\\
1636$+$351& KUV 16366+3506  &15.02&  36599& 7.99&  39212& $\pm$&10200&  36590& $\pm$& 4950& 0.66&  125.8\\
1647$+$591& G226-29      &12.23&  12460& 8.29&  12223& $\pm$&  400&  12272& $\pm$&  150& 0.79&   11.2\\
1650$+$724& HS 1650$+$7229 & --- &  44334& 7.52&  40820& $\pm$&25500&       &  --- &     & 0.49&  588.6\\
1657$+$343& PG 1657$+$344  &16.42&  52488& 7.62&  42918& $\pm$&19000&  41088& $\pm$& 6275& 0.54&  398.8\\
1713$+$695& G240-51      &13.27&  15241& 7.86&  14908& $\pm$& 1500&  15548& $\pm$&  825& 0.54&   28.3\\
1725$+$586& PG 1725$+$587  &15.70&  56084& 8.31&  49639& $\pm$&24000&  50541& $\pm$& 7800& 0.87&  168.0\\
1738$+$669& RE J1737$+$665 &14.60&  81210& 7.77&  56962& $\pm$&20000&  64926& $\pm$&13700& 0.67&  193.7\\
1749$+$717& HS 1749$+$7145 &15.70&  69363& 7.53&  36931& $\pm$&12000&  34196& $\pm$& 2750& 0.56&  365.3\\
1800$+$685& KUV 18004$+$6836&14.72&  44176& 7.82&  47201& $\pm$&16000&  42105& $\pm$& 8600& 0.60&  140.5\\
1819$+$580& RE J1820$+$580 &13.93&  43634& 7.93&  45125& $\pm$&16000&  44704& $\pm$& 9500& 0.64&   88.9\\
1827$+$778& HS 1827$+$7753 & --- &  74351& 7.50&  48129& $\pm$&22000&      &   --- &     & 0.57&  410.7\\
1828$+$668& KUV 18824$+$6650&16.65&  10798& 8.20&  10936& $\pm$& 4000&  10216& $\pm$& 1100& 0.73&   77.7\\
1845$+$019& Lanning 18     &12.95&  29384& 7.81&  28478& $\pm$& 2800&  28621& $\pm$&  900& 0.55&   45.2\\
1845$+$683& KUV 18453$+$6819&15.50&  37084& 8.20&  37563& $\pm$&10000&  43338& $\pm$& 5500& 0.77&  134.9\\
1855$+$338& G207-9       &14.63&  11958& 8.36&  11952& $\pm$&  600&  12055& $\pm$&  400& 0.83&   30.9\\
1919$+$145& GD 219         &12.98&  14430& 8.06&  14521& $\pm$&  800&  14935& $\pm$&  450& 0.65&   20.7\\
1935$+$276& G185-32      &13.03&  12123& 8.06&  12024& $\pm$&  450&  12039& $\pm$&  275& 0.64&   18.6\\
1936$+$327& GD 222         &13.58&  21329& 7.91&  20679& $\pm$& 2800&  21153& $\pm$&  700& 0.58&   41.8\\
1950$-$432& MCT 1950$-$4314&14.86&  39424& 7.86&  34706& $\pm$& 7000&  31565& $\pm$& 2125& 0.60&  135.1\\
1953$-$011& L997-21      &13.69&   7772& 8.24&   7150& $\pm$& 1000&   7637& $\pm$&  625& 0.75&   11.0\\
2007$-$303& LTT 7987       &12.18&  14454& 7.86&  14660& $\pm$&  800&  14913& $\pm$&  400& 0.54&   16.5\\
2014$-$575& L210-114     &13.70&  27465& 7.94&  26761& $\pm$& 2200&  26231& $\pm$&  725& 0.61&   54.3\\
2020$-$425& MCT 2020$-$4234&14.90&  29165& 8.10&  31689& $\pm$&10400&  27170& $\pm$& 1275& 0.70&   89.3\\
2028$+$390& GD 391         &13.38&  24127& 7.90&  24398& $\pm$& 1600&  24132& $\pm$&  450& 0.58&   42.6\\
2032$+$248& Wolf 1346      &11.52&  19953& 7.90&  19936& $\pm$& 1200&  19991& $\pm$&  450& 0.57&   15.4\\
2039$-$202& L711-10      &12.33&  19188& 7.93&  18575& $\pm$& 1200&  19169& $\pm$&  400& 0.58&   21.2\\
2046$+$396& KPD 2046$+$3940&14.43&  65428& 7.51&  50532& $\pm$&22000&  50801& $\pm$& 5800& 0.54&  200.9\\
2047$+$372& G210-36      &12.93&  14069& 8.21&  14064& $\pm$&  600&  14291& $\pm$&  350& 0.74&   17.9\\
2105$-$820& BPM 1266       &13.62&  10794& 8.19&  10385& $\pm$& 1300&  10674& $\pm$&  750& 0.72&   19.4\\
2111$+$498& GD 394         &13.08&  39205& 7.81&  32788& $\pm$& 3800&  34750& $\pm$& 2575& 0.58&   61.6\\
2117$+$539& G231-40      &12.26&  13991& 7.78&  14002& $\pm$& 1200&  14360& $\pm$&  550& 0.49&   17.5\\
2126$+$734& GW $+$73 8031  &12.83&  15287& 7.84&  15012& $\pm$&  700&  15393& $\pm$&  400& 0.53&   23.5\\
2136$+$828& G261-45      &13.02&  16905& 7.86&  16782& $\pm$&  800&  17488& $\pm$&  500& 0.55&   27.4\\
2146$-$433& MCT 2146$-$4320&15.81&  63711& 7.53& 121884& $\pm$&96000&  21596& $\pm$& 1825& 0.55&  364.7\\
2149$+$021& G93-48       &12.74&  17360& 7.93&  16801& $\pm$& 1000&  17248& $\pm$&  350& 0.58&   23.6\\
2153$-$419& MCT 2153$-$4156&15.38&  40974& 8.01&  27792& $\pm$&13000&  21549& $\pm$& 2900& 0.68&  157.2\\
2159$-$414& MCT 2159$-$4129&15.88&  61277& 7.48&  44506& $\pm$&18000&  54528& $\pm$&11100& 0.52&  386.2\\
2207$-$303& RE J2210$-$300 & --- &  28245& 7.86&  27871& $\pm$& 3600&       &  --- &     & 0.57&   74.3\\
2246$+$223& G67-23       &14.35&  10647& 8.80&  10161& $\pm$& 1000&  10559& $\pm$&  450& 1.10&   16.1\\
2309$+$105& GD 246         &13.09&  54681& 7.94&  51498& $\pm$&11800&  50217& $\pm$& 3100& 0.68&   67.3\\
2326$+$049& G29-38       &13.06&  11817& 8.15&  11632& $\pm$&  400&  11626& $\pm$&  175& 0.70&   17.3\\
2331$-$475& MCT 2331$-$473 &13.46&  52574& 7.78&  50344& $\pm$&15200&  48697& $\pm$& 4500& 0.60&   89.1\\
2341$+$322& L1512-34B    &12.90&  12573& 7.93&  12339& $\pm$&  350&  12420& $\pm$&  250& 0.57&   19.5\\
2349$+$286& PG 2349$+$286  &16.26&  37606& 7.99&  37906& $\pm$&17500&  31465& $\pm$& 6975& 0.66&  226.1\\
2353$+$026& PG 2353$+$026  &15.83&  60943& 7.60&  51961& $\pm$&25000&  33591& $\pm$& 3400& 0.56&  336.2\\
2357$+$296& PG 2357$+$297  &15.10&  52022& 7.53&  37553& $\pm$& 7000&  44920& $\pm$& 5500& 0.51&  233.3\\
2359$-$434& L362-81      &12.95&   8544& 8.44&   7832& $\pm$& 1000&   8148& $\pm$&  300& 0.88&    8.4\\
\enddata
\end{deluxetable}